\documentclass[aps,pra,amsmath,amssymb,showpacs]{revtex4}
\usepackage{graphicx,epsfig,epsf}
\usepackage{dcolumn}
\usepackage{bm}
\usepackage{mathbbol}
\usepackage{amssymb}
\usepackage{color}
\usepackage{graphicx}
\usepackage{subfigure}
\usepackage[english]{babel}

\usepackage{amsmath}
\usepackage{amsthm}
\usepackage{bbm}
 
\begin{document}
\newcommand{\ri}{ i}
\newcommand{\re}{ e}
\newcommand{\bx}{{\bf x}}
\newcommand{\bd}{{\bf d}}
\newcommand{\be}{{\bf e}}
\newcommand{\br}{{\bf r}}
\newcommand{\bk}{{\bf k}}
\newcommand{\bE}{{\bf E}}
\newcommand{\bI}{{\bf I}}
\newcommand{\bR}{{\bf R}}
\newcommand{\bZero}{{\bf 0}}
\newcommand{\bM}{{\bf M}}
\newcommand{\bn}{{\bf n}}
\newcommand{\bs}{{\bf s}}
\newcommand{\tbs}{\tilde{\bf s}}
\newcommand{\rSi}{{\rm Si}}
\newcommand{\beps}{\mbox{\boldmath{$\epsilon$}}}
\newcommand{\rg}{{\rm g}}
\newcommand{\tr}{{\rm tr}}
\newcommand{\xmax}{x_{\rm max}}
\newcommand{\ra}{{\rm a}}
\newcommand{\rx}{{\rm x}}
\newcommand{\rs}{{\rm s}}
\newcommand{\rP}{{\rm P}}
\newcommand{\up}{\uparrow}
\newcommand{\down}{\downarrow}
\newcommand{\hc}{H_{\rm cond}}
\newcommand{\kb}{k_{\rm B}}
\newcommand{\cI}{{\cal I}}
\newcommand{\tit}{\tilde{t}}
\newcommand{\cE}{{\cal E}}
\newcommand{\cC}{{\cal C}}
\newcommand{\Ubs}{U_{\rm BS}}
\newcommand{\qq}{{\bf ???}}
\newcommand*{\etal}{\textit{et al.}}
\def\vec#1{\mathbf{#1}}
\def\ket#1{|#1\rangle}
\def\bra#1{\langle#1|}
\def\keps{\mathbf{k}\boldsymbol{\varepsilon}}
\def\dm{\boldsymbol{\wp}}
\def\rev#1{{\bf #1}}

\newcommand{\Tr}{{\rm Tr}}

\newtheorem{definition}{Definition}
\newtheorem{proposition}{Proposition}
\newtheorem{theorem}{Theorem}
\newtheorem{remark}{Remark}
\newtheorem{corollary}{Corollary}
\newtheorem{lemma}{Lemma}
\newtheorem{example}{Example}

\sloppy

\title{Precision Measurements of Temperature and Chemical Potential
  of Quantum Gases}
\author{Ugo Marzolino$^{1,2,4}$ and Daniel Braun$^{3,4}$}
\affiliation{$^{1}$ Albert-Ludwigs-Universit\"at Freiburg, Hermann-Herder-Str. 3, D-79104 Freiburg, Deutschland\\
$^{2}$ Univerza v Ljubljani, Jadranska 19, SI-1000 Ljubljana, Slovenija}
\affiliation{$^3$ Universit\'e de Toulouse; UPS; Laboratoire de
 Physique Th\'eorique (IRSAMC); F-31062 Toulouse, France\\
$^4$ CNRS; LPT (IRSAMC); F-31062 Toulouse, France\\\\
}

\begin{abstract}
We investigate the sensitivity with which the temperature and the chemical
potential characterizing quantum gases can be measured.  We calculate
the corresponding quantum Fisher information matrices for both
fermionic and bosonic gases. For the latter, particular attention is
devoted to the situation close to the Bose-Einstein condensation
transition, which we examine not only for the standard scenario in
three dimensions, but also for generalized condensation in lower
dimensions, where the bosons condense in a subspace of Hilbert space
instead of a unique ground state, as well as condensation at fixed volume or
fixed pressure.  We show that Bose Einstein condensation can lead to sub-shot noise sensitivity for the measurement of the chemical potential. We also examine the influence of interactions on the sensitivity in three different models, and show that mean-field and contact interactions deteriorate the sensitivity but only slightly for experimentally accessible weak interactions.
\end{abstract}


\maketitle

\section{Introduction}
The use of gases for measurements of temperature and pressure has a long
history, going back at least as far as Galileo Galilei in the 16th
century,  who built a  
thermoscope based on a glass pipe filled with air and sealed with a water
surface.  Variations in
temperature show up as variations of the level of the water, as
depending on the contraction of the air water gets   
sucked up to different levels \cite{Benedict84}. Still today, gas thermometry
plays an important role for the calibration of other thermometers,
even though a lot of effort has to be spent to compensate for many
effects of real working substances \cite{Schooley90}. A high precision
solid state 
thermometer for temperatures spanning four orders of magnitude based
on the noise properties of an electron gas was developed by Spietz et
al.~\cite{Spietz03}. Shot-noise thermometry of an electron gas was
also applied recently to quantum
Hall edge states \cite{Levkivskyi12}. 

The chemical potential is, besides temperature, the only other independent parameter that characterizes ideal quantum gases, and one may legitimately
ask how sensitively both parameters can be measured in principle. For
charged gases, such as the electron gas responsible for electrical
conductance in a metal or semi-conductor, the chemical potential is directly
linked to voltage.  More precisely, without current flow, the voltage drop
between two parts of a sample is given by the difference between the
chemical potentials divided by the electron charge, $V=(\mu_2-\mu_1)/e$,
such that the precision of a measurement of the chemical potential
translates directly into the precision of voltage measurement
\cite{Madelung08}. Voltage measurements on the other hand are at the basis
of a huge variety of modern sensors, such that knowing the ultimate
precision with which chemical potentials can be measured is of fundamental
importance.
For instance, voltage measurements can be employed in magnetometry in the presence of Hall effects or magnetoresistances, as an alternative to other metrological schemes \cite{Savukov2005,Fang2013,Loretz2013,Aiello2013,Steinke,Eto,Waxman,Puentes}.

Establishing the ultimate bounds on sensitivity of measurements is
one of the major goals of parameter estimation theory. This theoretical
frame work was developed in classical statistical analysis \cite{Cramer46},
and later generalized to the quantum world
\cite{Holevo,Helstrom,Braunstein94}.  It leads to the (quantum) Cram\'er-Rao 
bound that establishes that under suitable regularity conditions and
for unbiased estimators the best 
sensitivity with which a parameter $x$ can be 
measured is given by the inverse of the (quantum) Fisher information. In
classical statistical analysis, the 
classical Fisher information 
characterizes the probability distribution
of the measurement results $A_i$ of the measured quantity $A$ (which may be
different from $x$) given the parameter $x$.  The corresponding bound is
optimized over all 
estimator functions.  In the quantum world, all statistical information is
coded in the quantum state (density matrix) of the system characterized by
the parameter $x$, and the quantum Cram\'er-Rao bound is obtained
from the 
classical Cram\'er-Rao bound by additionally optimizing over all possible
positive operator-valued measure (POVM) measurements. For unbiased
estimators the bound is tight. As a result, it
represents the best sensitivity with which a parameter can be measured, no
matter what the measurement strategy, data analysis, feedback schemes etc.  \\


In the quantum physics community interest has
recently arisen in the question whether quantum mechanical effects may be
used for enabling or improving temperature measurements
\cite{Stace10,Ruostekoski09,Sabin13}. In \cite{Stace10} it was shown that
for any thermalizing thermometer with extensive internal energy the
sensitivity scales  as $1/\sqrt{N}$, 
corresponding to a linear scaling of the quantum Fisher information with
$N$. This scaling is called shot-noise limit or standard quantum limit
(SQL). On 
the other hand, it is well-known for the measurement of other quantities
that using quantum effects one can beat in principle the 
SQL. Examples include the use of squeezed states \cite{Caves81} (which 
allow one to change the prefactor of the linear scaling of the quantum
Fisher information with $N$, a strategy
implemented in Advanced-LIGO), or 
entangled states \cite{Giovannetti04,Giovannetti06,Giovannetti11}, which can
enable reaching the so-called Heisenberg limit (HL) characterized by a Fisher
information that scales as $N^2$. A popular
entangled state is the NOON state \cite{Boto00}. It was shown in
\cite{Stace10} that the principles of 
interferometric quantum-enhanced metrology can be applied to the
measurement  of temperature, allowing one at least in 
principle to achieve HL scaling. In several other systems the
sensitivity of quantum mechanical interference to thermal noise was
already used for thermometry \cite{Brunelli11,Slodicka12}, but no
attempts were made for establishing the ultimate sensitivity of such
an approach.\\

Few 
experiments using entangled states have surpassed the SQL, and all
experiments have been limited to small values of $N$, due to the extreme
sensitivity of such states to decoherence. Indeed it has become clear that
the smallest amount of Markovian decoherence leads back to the SQL
\cite{Huelga97,Kolodynski10,Escher2011,Alipour2013}, limiting the method to
short time analyses \cite{Benatti2013-2}  or niche applications
\cite{Gross2010,Riedel2010,Wolfgramm13}.  
On the other hand, while an entangled state of distinguishable
  particles is needed for beating the SQL in interferometric metrology, a
  HL-like scaling can be reached by feeding interferometers with unentangled
  states of identical particles that cannot be distinguished by any degree
  of freedom
  \cite{Holland1993,Benatti2010,Benatti2011,Argentieri2011,Benatti2013}. Another
  strategy is the use of interactions between the $N$ particles 
\cite{Luis04,Luis07,Napolitano11}. With $k$-body interactions one may even
surpass the 
HL, but interactions between all particles are required, which makes such
models unphysical for large particle numbers due to the resulting
non-extensive 
character of the total energy. In \cite{Braun11} a method of ``coherent
averaging'' was proposed for reaching the HL, in which the $N$ constituents
interact with a 
common ``quantum bus'' which is then read out. The quantum bus can even be
an environment of which one has no full control, leading to the possibility
of using collective decoherence effects for precision measurements
\cite{Braun09}.  \\

In the present paper, we shall discuss the scaling
of the Fisher matrix characterizing measurements of both temperature
and chemical potential of ideal quantum gases with the average number
of particles, as well as of three different models of interacting
particles, which will establish the ultimate sensitivity with which
these parameters can be measured. We pay particular attention to the
influence of the Bose-Einstein condensation (BEC) phase 
transition in bosonic gases.  Indeed, it is well
known that phase transitions can lead to enhanced susceptibilities, as is
witnessed by large fluctuations \cite{Huang}, and, closely 
related, to large quantum Fisher information. Discontinuities
in the quantum Fisher information were proposed before as a tool for
detecting phase transitions in the absence of knowledge of an order
parameter \cite{Zanardi06}. We will see that indeed the onset of BEC can
improve the sensitivity of a measurement of the chemical
potential beyond the shot-noise limit. We carefully analyse several
scenarios of BEC: the standard case of fixed density in the
thermodynamic limit, the case of cooling at fixed volume, the case
of isobaric cooling, as well as generalized BEC in lower dimensions,
where condensation occurs in a subspace of Hilbert space, instead of
only the ground state.  We also examine the influence of interactions
on the sensititivy in the framework of three different models, and
show that interactions can be detrimental for the sensitivity with
which the chemical potential can be measured.

The paper is organized as follows. In sections \ref{sec.metrology} and
\ref{sec.gases} preliminary discussions respectively on quantum metrology
and quantum gases are presented. Our results, based on these preliminary
notions, are reported in section \ref{sec.cont} for gases in the continuum
approximation, in section \ref{sec.BEC} for BEC, and in section
\ref{sec.int} for bosonic interactiong gases. Conclusions are discussed in
section \ref{sec.concl}, and some technical details in the appendices.

\section{Metrology with quantum gases} \label{sec.metrology}
We start by introducing the formalism of parameter estimation in the
context  of quantum gases. Consider the hamiltonian $H$ and the total number of particles is described by the operator 
\begin{equation} \label{numb}
N=\sum_k a_k^\dag a_k,
\end{equation}

\noindent
where $a_k^\dag$ and $a_k$ are the creation and annihilation operators of the $k$-th mode. In condensed matter systems, the hamiltonian and the total number of particles commute with each other, and the common eigenvectors are $|N,E_N\rangle$ with particle number eigenvalues $N$ and hamiltonian eigenvalues $E_N$. The grand canonical thermal state is 

\begin{equation} \label{grandcan.gen}
\rho_{\beta,\mu}=\frac{e^{-\beta(H-\mu N)}}{Z_G}=\sum_{N,E_N}\rho_{\beta,\mu}^{(N,E_N)}|N,E_N\rangle\langle N,E_N|, \qquad \rho_{\beta,\mu}^{(N,E_N)}=\frac{e^{-\beta(E_N-\mu N)}}{Z_G}
\end{equation}

\noindent
where we have defined the grand canonical partition function $Z_G=\sum_{N,E_N}\rho_{\beta,\mu}^{(N,E_N)}$.

In statistical mechanics, the inverse temperature $\beta$ and the chemical potential $\mu$ are the
Lagrange multipliers of the average energy and the average total number of particle respectively. These two latter quantities fix $(\beta,\mu)$. Therefore, one way of estimating $(\beta,\mu)$ is through measuring  the average energy and the  number of particles. We will discuss how this is related to the best sensitivity for the estimation of $(\beta,\mu)$. 

Quantum estimation theory \cite{Helstrom,Holevo} provides a bound for the
covariance matrix of the parameter estimators. The ingredients are the
symmetric logarithmic derivatives with respect to the parameters
$\lambda_1=\beta$, $\lambda_2=\mu$ 
\begin{equation} \label{sld.gen}
L_{\lambda_j}=\sum_{N,E_N}\frac{\partial_{\lambda_j}\rho_{\beta,\mu}^{(N,E_N)}}{\rho_{\beta,\mu}^{(N,E_N)}}|N,E_N\rangle\langle N,E_N|=
\begin{cases}
\mu N-H-\langle\mu N-H\rangle & \mbox{if } j=1 \\
\beta\left(N-\langle N\rangle\right) & \mbox{if } j=2
\end{cases},
\end{equation}

\noindent
and the quantum Fisher matrix has entries

\begin{equation} \label{fisher.gen}
F_{\lambda_j,\lambda_l}=\frac{1}{2}{\rm tr}\left(\rho\{L_{\lambda_j},L_{\lambda_l}\}\right)=\sum_{N,E_N}\frac{(\partial_{\lambda_j}\rho_{\beta,\mu}^{(N,E_N)})(\partial_{\lambda_l}\rho_{\beta,\mu}^{(N,E_N)})}{\rho_{\beta,\mu}^{(N,E_N)}}.
\end{equation}

\noindent
where $\{,\}$ denotes the anti commutator. The covariance matrix of any estimator of $\lambda$ is bounded by the
quantum Cram\'er-Rao bound,

\begin{equation} \label{CRB}
\begin{pmatrix}
\displaystyle {\rm var}(\beta) & {\rm cov}(\beta,\mu) \\
\displaystyle {\rm cov}(\beta,\mu) & {\rm var}(\mu)
\end{pmatrix}\geqslant F^{-1}=
\begin{pmatrix}
\displaystyle \frac{F_{\mu,\mu}}{F_{\mu,\mu}F_{\beta,\beta}-F_{\mu,\beta}^2} & \displaystyle \frac{F_{\mu,\beta}}{F_{\mu,\beta}^2-F_{\mu,\mu}F_{\beta,\beta}} \\
\displaystyle \frac{F_{\mu,\beta}}{F_{\mu,\beta}^2-F_{\mu,\mu}F_{\beta,\beta}} & \displaystyle \frac{F_{\beta,\beta}}{F_{\mu,\mu}F_{\beta,\beta}-F_{\mu,\beta}^2}
\end{pmatrix},
\end{equation}
where ${\rm var}$ and ${\rm cov}$ are the variances and the covariance of
the estimation problem, and $A\geqslant B$ means that $A-B$ is a semi-positive
definite matrix. 

If one parameter, say $\beta$ ($\mu$), is known the inverse of the diagonal
entry $F_{\mu,\mu}$ ($F_{\beta,\beta}$) is the best sensitivity for the
estimation of $\mu$ ($\beta$). A non-diagonal Fisher matrix,
i.e. $F_{\mu,\beta}\neq 0$, means that estimation of $\beta$ and $\mu$ are
correlated. Thus, the diagonal entries of the inverse Fisher matrix $F^{-1}$
are the optimal sensitivities of each parameter in a joint measurement, and 
the off-diagonal term is the corresponding covariance.

Denoting with $\Delta^2$ and ${\rm Cov}$ respectively the variance and the covariance in the grand canonical state, the Fisher matrix explicitly reads

\begin{eqnarray}
\label{entries1.gen} & & F_{\mu,\mu}=\beta^2\Delta^2 N, \\
\label{entries2.gen}& & F_{\beta,\beta}=\Delta^2 (\mu N-H), \\
\label{entries3.gen}& & F_{\mu,\beta}=F_{\beta,\mu}=\beta \, {\rm Cov}(N,\mu N-H),
\end{eqnarray}

\noindent
The latter equations can be computed by the standard relations

\begin{eqnarray}
\label{rel1} & & \Delta^2 N=\frac{1}{\beta^2}\frac{\partial^2}{\partial\mu^2}\ln Z_G=\frac{1}{\beta}\frac{\partial}{\partial\mu}\langle N\rangle, \\
\label{rel2} & & \Delta^2(\mu N-H)=\frac{\partial^2}{\partial\beta^2}\ln Z_G=\frac{\partial}{\partial\beta}\langle \mu N-H\rangle, \\
\label{rel3} & & {\rm Cov}(N,\nu N-H)=\frac{\partial}{\partial\beta}\frac{1}{\beta}\frac{\partial}{\partial\mu}\ln Z_G=\frac{\partial}{\partial\beta}\langle N\rangle,
\end{eqnarray}

The optimal measurement is a projective measurement onto
the eigenstates of its symmetric logarithmic derivatives
\cite{Helstrom,Holevo}. For the grand canonical thermal states considered
here, the symmetric logarithmic derivatives commute with each other. Hence,
the two parameters can be simultaneously measured, contrary to the general
case of multivariate quantum metrology. Our problem is a very special case
of quantum estimation where only the eigenvalues of the state depend on the
parameters $(\beta,\mu)$, and the estimation problem becomes a classical
problem in the representation of the Fock basis (\ref{fock}).

For a temperature measurement which is known to be difficult, different
approaches have been proposed. One is based on the measurement
of a spin gradient between two domains of a spin mixture separated by a
magnetic field gradient \cite{Weld2009}. In \cite{Leanhardt2013}, where a temperature of
500\,$p$K was reached, temperature was calibrated to the BEC transition
temperature by measuring trap frequency and average particle number.  Zhou
and Ho proposed the use of local particle number fluctuations which are
related to temperature through a generalized dissipation fluctuation theorem
\cite{Zhou2011}.  Temperature measurements based on fluctuations of the
total particle number were realized in \cite{Muller2010,Sanner2010}. Our
results show that the measurement of the particle number fluctuations
is optimal for the estimation of  chemical potential.  Particle counting has 
been implemented for fermionic \cite{Muller2010,Sanner2010} and bosonic
\cite{Greiner2001,Esteve2006,Bouchoule2011,Armijo2011} gases, but, up to our
knowledge, measurement of the chemical potential based on this technique has not been implemented yet.

Some general remarks are in order. The variance of the number of particles
is directly connected to the thermodynamic stability, through the isothermal
compressibility $\kappa_T$ \cite{Huang}, which measures how the system
responds to variations of the pressure: 

\begin{equation} \label{kappa}
\kappa_T=-\frac{1}{V}\left(\frac{\partial V}{\partial P}\right)_T=\frac{\beta\Delta^2 N}{\varrho \, \langle N\rangle}.
\end{equation}

\noindent
Hence, any superlinear scaling of the Fisher information $F_{\mu,\mu}$
implies thermodynamical instability via (\ref{entries1.gen}), i.e. the
compressibility grows with the number of particles, diverges in the
thermodynamic limit \cite{Huang}, and ceases to be an intensive quantity. The
thermodynamic instability of 
superlinear particle number fluctuations in the grand canonical state has
been used for claiming that the inappropriate application
of the grand canonical ensemble may give rise to unphysical
results \cite{Yukalov2005}. But the grand canonical thermal is a
physically meaningful state for the following reasons: Firstly, the grand 
canonical state naturally arises as the equilibrium state when particles can
be exchanged with the thermal bath \cite{Huang,Reichl}. Secondly, it has
been argued that the unique correct definition of the chemical potential for
finite systems is provided by the grand canonical state, even if there are
other inequivalent definitions converging to the same quantity in the
thermodynamic limit \cite{Kaplan2006}. Finally, superlinear fluctuations can
be stabilized and observed within mesoscopic sizes
\cite{Greiner2001,Esteve2006,Bouchoule2011,Armijo2011}. For these reasons we
will base our analysis on the grand-canonical ensemble.

\subsection{Optimal measurements}
\label{joint.meas}

We now derive the joint measurement of the chemical potential and the inverse temperature that attains the quantum Cram\'er-Rao bound \eqref{CRB}. As we mentioned, the two symmetric logarithmic derivatives commute with each other. This allows the unusual situation where the two parameters $(\beta,\mu)$ can be measured simultaneously. In this case, applying the transformation which diagonalizes the matrix $F^{-1}$ to the inequality \eqref{CRB}, we find two uncorrelated estimations, i.e. with vanishing covariance, of linear combinations of the parameters $(\beta,\mu)$. At this aim, it is convenient to consider dimensionless parameters which can be summed without issues about physical dimensions. The dimensionless parameters are $(\bar\beta=\beta/\beta_0,\bar\mu=\mu/\mu_0)$, where $\beta_0$ and $\mu_0$ are constant values. Examples for the case of fixed volume are $\mu_0=\beta_0^{-1}=2\pi^2\hslash^2/(mV_d^{2/d})$ for homogeneous gases and $\mu_0=\beta_0^{-1}=\hslash\Omega_d$ for harmonically trapped gases; if the density is fixed, one can choose $\mu_0=\beta_0^{-1}=2\pi^2\hslash^2\rho^{2/d}/m$ for homogeneous gases and $\mu_0=\beta_0^{-1}=\hslash\tilde\rho^{1/d}$ for harmonically trapped gases. The quantum Cram\'er-Rao bound for the dimensionless parameters is

\begin{equation} \label{CRB2}
\begin{pmatrix}
\displaystyle {\rm var}(\bar\beta) & {\rm cov}(\bar\beta,\bar\mu) \\
\displaystyle {\rm cov}(\bar\beta,\bar\mu) & {\rm var}(\bar\mu)
\end{pmatrix}\geqslant
\begin{pmatrix}
\displaystyle \frac{F_{\bar\mu,\bar\mu}}{F_{\bar\mu,\bar\mu}F_{\bar\beta,\bar\beta}-F_{\bar\mu,\bar\beta}^2} & \displaystyle \frac{F_{\bar\mu,\bar\beta}}{F_{\bar\mu,\bar\beta}^2-F_{\bar\mu,\bar\mu}F_{\bar\beta,\bar\beta}} \\
\displaystyle \frac{F_{\bar\mu,\bar\beta}}{F_{\bar\mu,\bar\beta}^2-F_{\bar\mu,\bar\mu}F_{\bar\beta,\bar\beta}} & \displaystyle \frac{F_{\bar\beta,\bar\beta}}{F_{\bar\mu,\bar\mu}F_{\bar\beta,\bar\beta}-F_{\bar\mu,\bar\beta}^2}
\end{pmatrix},
\end{equation}
with

\begin{equation}
F_{\bar\beta,\bar\beta}=\beta_0^2 F_{\beta,\beta}, \qquad F_{\bar\mu,\bar\mu}=\mu_0^2 F_{\mu,\mu}, \qquad F_{\bar\beta,\bar\mu}=\beta_0\mu_0 F_{\beta,\mu}.
\end{equation}
The right-hand-side of the inequality \eqref{CRB2} is diagonalized by the following orthogonal matrix

\begin{equation}
R=
\begin{pmatrix}
\displaystyle \frac{F_{\bar\beta,\bar\beta}-F_{\bar\mu,\bar\mu}-\sqrt{4F_{\bar\beta,\bar\mu}^2+(F_{\bar\beta,\bar\beta}-F_{\bar\mu,\bar\mu})^2}}{\sqrt{4F_{\bar\beta,\bar\mu}^2+\left(F_{\bar\beta,\bar\beta}-F_{\bar\mu,\bar\mu}-\sqrt{4F_{\bar\beta,\bar\mu}^2+(F_{\bar\beta,\bar\beta}-F_{\bar\mu,\bar\mu})^2}\right)^2}}
& \displaystyle \frac{2 F_{\bar\beta,\bar\mu}}{\sqrt{4F_{\bar\beta,\bar\mu}^2+\left(F_{\bar\beta,\bar\beta}-F_{\bar\mu,\bar\mu}-\sqrt{4F_{\bar\beta,\bar\mu}^2+(F_{\bar\beta,\bar\beta}-F_{\bar\mu,\bar\mu})^2}\right)^2}} \\
\displaystyle \frac{F_{\bar\beta,\bar\beta}-F_{\bar\mu,\bar\mu}+\sqrt{4F_{\bar\beta,\bar\mu}^2+(F_{\bar\mu,\bar\mu}-F_{\bar\beta,\bar\beta})^2}}{\sqrt{4F_{\bar\beta,\bar\mu}^2+\left(F_{\bar\beta,\bar\beta}-F_{\bar\mu,\bar\mu}+\sqrt{4F_{\bar\beta,\bar\mu}^2+(F_{\bar\beta,\bar\beta}-F_{\bar\mu,\bar\mu})^2}\right)^2}}
& \displaystyle \frac{2 F_{\bar\beta,\bar\mu}}{\sqrt{4F_{\bar\beta,\bar\mu}^2+\left(F_{\bar\beta,\bar\beta}-F_{\bar\mu,\bar\mu}+\sqrt{4F_{\bar\beta,\bar\mu}^2+(F_{\bar\beta,\bar\beta}-F_{\bar\mu,\bar\mu})^2}\right)^2}} \\
\end{pmatrix}.
\end{equation}
The quantum Cram\'er-Rao bound \eqref{CRB2} is then transformed into

\begin{equation} \label{CRB3}
\begin{pmatrix}
\displaystyle {\rm var}(\bar\lambda_1) & {\rm cov}(\bar\lambda_1,\bar\lambda_2) \\
\displaystyle {\rm cov}(\bar\lambda_1,\bar\lambda_2) & {\rm var}(\bar\lambda_2)
\end{pmatrix}=R\begin{pmatrix}
\displaystyle {\rm var}(\bar\beta) & {\rm cov}(\bar\beta,\bar\mu) \\
\displaystyle {\rm cov}(\bar\beta,\bar\mu) & {\rm var}(\bar\mu)
\end{pmatrix}R^T \nonumber\geqslant
\begin{pmatrix}
\displaystyle F_{\bar\lambda_1}^{-1} & 0 \\
0 & \displaystyle F_{\bar\lambda_2}^{-1}
\end{pmatrix},
\end{equation}
with

\begin{equation}
\begin{pmatrix}
\displaystyle \bar\lambda_1 \\
\displaystyle \bar\lambda_2
\end{pmatrix}=R\begin{pmatrix}
\displaystyle \bar\beta \\
\displaystyle \bar\mu
\end{pmatrix}, \qquad F_{\bar\lambda_j}=\frac{1}{2}\left(F_{\bar\beta,\bar\beta}+F_{\bar\mu,\bar\mu}+(-1)^j\sqrt{4F_{\bar\beta,\bar\mu}^2+(F_{\bar\beta,\bar\beta}-F_{\bar\mu,\bar\mu})^2}\right).
\end{equation}

The symmetric logarithmic derivatives with respect to the parameters $\bar\lambda_{1,2}$ are

\begin{equation} \label{sld.gen1}
L_{\bar\lambda_j}=\sum_{N,E_N}\frac{\partial_{\bar\lambda_j}\rho_{\beta,\mu}^{(N,E_N)}}{\rho_{\beta,\mu}^{(N,E_N)}}|N,E_N\rangle\langle N,E_N|,
\end{equation}
and explicitely

\begin{equation}
\begin{pmatrix}
L_{\bar\lambda_1} \\
L_{\bar\lambda_2}
\end{pmatrix}=R
\begin{pmatrix}
\beta_0 & 0 \\
0 & \mu_0
\end{pmatrix}
\begin{pmatrix}
L_{\beta} \\
L_{\mu}
\end{pmatrix}=R
\begin{pmatrix}
\beta_0 & 0 \\
0 & \mu_0
\end{pmatrix}
\begin{pmatrix}
\mu N-H-\langle\mu N-H\rangle \\
\beta\left(N-\langle N\rangle\right)
\end{pmatrix}.
\end{equation}
According to  quantum estimation theory \cite{Holevo,Helstrom}, the optimal estimation is then given by measuring the following observables

\begin{equation}
\begin{pmatrix}
O_{\bar\lambda_1} \\
O_{\bar\lambda_2}
\end{pmatrix}=
\begin{pmatrix}
\displaystyle \bar\lambda_1 \mathbb{1}+\frac{L_{\bar\lambda_1}}{F_{\bar\lambda_1}} \\
\displaystyle \bar\lambda_2 \mathbb{1}+\frac{L_{\bar\lambda_2}}{F_{\bar\lambda_2}}
\end{pmatrix}=R
\begin{pmatrix}
\beta_0 & 0 \\
0 & \mu_0
\end{pmatrix}
\begin{pmatrix}
\displaystyle \frac{\beta}{\beta_0^2}\mathbb{1}+\frac{\mu N-H-\langle\mu N-H\rangle}{F_{\bar\lambda_1}} \\
\displaystyle \frac{\mu}{\mu_0^2}\mathbb{1}+\frac{\beta\left(N-\langle N\rangle\right)}{F_{\bar\lambda_2}}
\end{pmatrix}.
\end{equation}

For the single parameter estimation of $\beta$ or $\mu$, when the other parameter in known, the optimal estimations are given respectively by the measurement of following observables

\begin{eqnarray}
O_{\beta} & = & \beta\mathbb{1}+\frac{L_{\beta}}{F_{\beta,\beta}}=\beta\mathbb{1}+\frac{\mu N-H-\langle\mu N-H\rangle}{\Delta^2(\mu N-H)}, \\
O_{\mu} & = & \mu\mathbb{1}+\frac{L_{\mu}}{F_{\mu,\mu}}=\mu\mathbb{1}+\frac{N-\langle N\rangle}{\beta\Delta^2 N}.
\end{eqnarray}

It is straightforwad to show that the above estimators are unbiased and attain the quantum Cram\'er-Rao bound \eqref{CRB3}:

\begin{eqnarray}
&& \langle O_{\bar\lambda_j}\rangle=\bar\lambda_j, \qquad
\Delta^2 O_{\bar\lambda_j}=\frac{1}{F_{\bar\lambda_j}}, \qquad
{\rm Cov}(O_{\bar\lambda_1},O_{\bar\lambda_2})=0, \\
&& \langle O_{\beta}\rangle=\beta, \qquad
\Delta^2 O_{\beta}=\frac{1}{F_{\beta,\beta}}, \\
&& \langle O_{\mu}\rangle=\mu, \qquad
\Delta^2 O_{\mu}=\frac{1}{F_{\mu,\mu}}.
\end{eqnarray}
This optimal joint estimation
can be realized e.g.~by measuring the energy and the number of particles. In general, in quantum parameter estimation the optimal measurement depends of the parameters to be estimated, and thus is called \emph{local} estimation. Interestingly, the optimal sensitivities for single parameter estimation of $\mu$ and $\beta$, when the other parameter is known, are achieved by a \emph{global} estimation, in the sense that the optimal measurement can be implemented with an operator that does not depend on the parameter to be estimated. The optimal estimators themself do depend on the parameters to be estimated, but that prior knowledge is only needed on the level of the data analysis, not for the choice of the measurement itself. This unusual situation is reminiscent once more of the situation in classical estimation theory, where the measurement is fixed from the beginning, and can be tracked back to the exponential form of the density matrix as function of $\beta(H-\mu N)$, which makes that the logarithmic derivatives only depend on these operators, but not anymore on the corresponding parameter.

Another joint optimal estimation can be derived from the consideration that our estimation problem reduces to a classical problem in the joint eigenbasis of $H$ and of $N$. Consequently, the quantum Fisher information is the classical Fisher information of the probability distribution $\rho_{\beta,\mu}^{(N,E_N)}$. It is known that the maximum likelihood estimator is asymptotically biased and optimal, in the sense of achieving the classical Cram\'er-Rao bound, in the limit of infinitely many measurements \cite{Helstrom,Holevo}. The Cram\'er-Rao bound for $M$ measurements reads

\begin{equation} \label{CRBM}
\begin{pmatrix}
\displaystyle {\rm var}(\beta) & {\rm cov}(\beta,\mu) \\
\displaystyle {\rm cov}(\beta,\mu) & {\rm var}(\mu)
\end{pmatrix}\geqslant(MF)^{-1}.
\end{equation}
Consider the outcomes of particle number and energy measurements $\{N^{(i)},E^{(i)}_{N^{(i)}}\}_{i=1,\dots,M}$. The maximum likelihood esestimation consists in maximizing the average logarithmic likelihood ${\cal L}=\frac{1}{M}\ln\prod_i\rho_{\beta,\mu}^{(N^{(i)},E^{(i)}_{N^{(i)}})}$ with respect to the parameters to be estimated. This maximization problem is equivalent to impose vanishing derivatives

\begin{equation}
\begin{cases}
\displaystyle \frac{\partial{\cal L}}{\partial\beta}=\frac{1}{M}\sum_{i=1}^M\left(\mu N^{(i)}-E^{(j)}_{N^{(i)}}\right)-\langle\mu N-H\rangle=0, \\
\displaystyle \frac{\partial{\cal L}}{\partial\mu}=\frac{\beta}{M}\sum_{i=1}^M N^{(i)}-\beta\langle N\rangle=0,
\end{cases} \label{maxLH}
\end{equation}
provided a negative Hessian matrix

\begin{equation}
\left.
\begin{pmatrix}
\displaystyle \frac{\partial^2{\cal L}}{\partial\beta^2} & \displaystyle \frac{\partial^2{\cal L}}{\partial\beta\partial\mu} \\
\displaystyle \frac{\partial^2{\cal L}}{\partial\beta\partial\mu} & \displaystyle \frac{\partial^2{\cal L}}{\partial\mu^2}
\end{pmatrix}\right|_{\frac{\partial{\cal L}}{\partial\beta}=0,\frac{\partial{\cal L}}{\partial\mu}=0}
=
\begin{pmatrix}
F_{\beta,\beta} & F_{\beta,\mu} \\
F_{\beta,\mu} & F_{\mu,\mu}
\end{pmatrix}.
\end{equation}
Equations \eqref{maxLH} imply that the maximum likelihood estimator consists in finding the parameters $(\beta,\mu)$ for which the experimental averages $\frac{1}{M}\sum_{i=1}^M N^{(i)}$ and $\frac{1}{M}\sum_{i=1}^M E^{(i)}_{N^{(i)}}$ equal the theoretical quantities $\langle N\rangle$ and $\langle H\rangle$ respectively. The advantage of this estimator is that it does not require the a priori knowledge of the parameters $(\beta,\mu)$ even in the data analysis, although this property as well as the optimality hold true only in the limit $M\to\infty$.

Interestingly, the maximum likelihood estimator of single parameters, e.g. inverting $\frac{1}{M}\sum_{i=1}^M(\mu N^{(i)}-E^{(i)}_{N^{(i)}})=\langle\mu N-H\rangle$ for $\beta$ or $\frac{1}{M}\sum_{i=1}^M N^{(i)}=\langle N\rangle$ for $\mu$, achieves the Cram\'er-Rao bound \eqref{CRBM} for any finite $M$. Indeed, $\langle N\rangle$ and $\langle\mu N-H\rangle$ are estimated from a finite sample of i.i.d. couples $(N^{(i)},E^{(i)}_{N^{(i)}})$ by $\frac{1}{M}\sum_{i=1}^M N^{(i)}$ and $\frac{1}{M}\sum_{i=1}^M(\mu N^{(i)}-E^{(i)}_{N^{(i)}})$ with standard deviations $\langle(\frac{1}{M}\sum_{i=1}^M N^{(i)}-\langle N\rangle)^2\rangle=\frac{1}{M}\Delta^2N$ and $\langle(\frac{1}{M}\sum_{i=1}^M(\mu N^{(i)}-E^{(i)}_{N^{(i)}})-\langle\mu N-H\rangle)^2\rangle=\frac{1}{M}\Delta^2(\mu-H)$ respectively. The variance of the estimation of $\mu$ ($\beta$) can be computed through simple laws of error propagation from the measurement of $\langle N\rangle$ ($\langle\mu N-H\rangle$):

\begin{eqnarray}
{\rm var}(\mu) & = & \left(\frac{\partial\mu}{\partial\langle N\rangle}\right)^2\frac{\Delta^2N}{M}=\frac{1}{M\beta^2\Delta^2N}, \\
{\rm var}(\beta) & = & \left(\frac{\partial\beta}{\partial\langle\mu N-H\rangle}\right)^2\frac{\Delta^2(\mu N-H)}{M}=\frac{1}{M\Delta^2(\mu N-H)}.
\end{eqnarray}


\subsection{Ideal gases} \label{sec.ideal}

We now focus on ideal gases of fermions or bosons. The hamiltonian in second quantization is
\begin{equation} \label{ham}
H=\sum_k \varepsilon_k a_k^\dag a_k,
\end{equation}

\noindent
where $\varepsilon_k$ is the energy of a single particle filling the $k$-th mode. The eigenvectors of both the hamiltonian and the total number of particles are the following Fock states

\begin{equation} \label{fock}
|\{n_k\}\rangle=\prod_k\frac{(a_k^\dag)^{n_k}}{\sqrt{n_k!}}|0\rangle=\bigotimes_k|n_k\rangle,
\end{equation}

\noindent
where $|0\rangle=\bigotimes_k|0_k\rangle$ is the vacuum, and we have used the tensor decomposition in terms of the single mode Fock states
$|n_k\rangle=(a_k^\dag)^{n_k}/\sqrt{n_k!}|0_k\rangle$. The grand canonical thermal state is 

\begin{equation} \label{grandcan}
\rho_{\beta,\mu}=\frac{e^{-\beta(H-\mu N)}}{Z_G}=\bigotimes_k\sum_{n_k}\rho_{\beta,\mu}^{(n_k)}|n_k\rangle\langle n_k|, \qquad \rho_{\beta,\mu}^{(n_k)}=\frac{e^{-\beta n_k(\varepsilon_k-\mu)}}{Z_k}
\end{equation}

\noindent
where we have defined the grand canonical partition function $Z_G=\prod_k Z_k$

\begin{equation}
Z_k=
\begin{cases}
1+e^{-\beta(\varepsilon_k-\mu)} & \mbox{for fermions} \\
\displaystyle \frac{1}{1-e^{-\beta(\varepsilon_k-\mu)}} & \mbox{for bosons}
\end{cases}.
\end{equation}

\noindent
The sums over $n_k$ run from zero to one for fermions and from zero to
infinity for bosons. 

The average total number of particles and the average energy are, respectively,

\begin{eqnarray}
\label{N} & & \langle N\rangle=\sum_k\frac{1}{e^{\beta(\varepsilon_k-\mu)}\pm 1}, \\
\label{H} & & \langle H\rangle=\sum_k\frac{\varepsilon_k}{e^{\beta(\varepsilon_k-\mu)}\pm 1},
\end{eqnarray}

\noindent
where the plus signs  hold for fermions and the minus signs  for bosons.

The symmetric logarithmic derivatives with respect to the parameters
$\lambda_1=\beta$, $\lambda_2=\mu$ are
\begin{equation} \label{sld}
L_{\lambda_j}=\bigotimes_k\sum_{n_k}\left(\sum_{k'}\frac{\partial_{\lambda_j}\rho_{\beta,\mu}^{(n_{k'})}}{\rho_{\beta,\mu}^{(n_{k'})}}\right)|n_k\rangle\langle n_k|=
\begin{cases}
\mu N-H-\langle\mu N-H\rangle & \mbox{if } j=1 \\
\beta\left(N-\langle N\rangle\right) & \mbox{if } j=2
\end{cases},
\end{equation}

\noindent
and the quantum Fisher matrix $F=[F_{\lambda_j,\lambda_l}]$ with entries

\begin{equation} \label{fisher}
F_{\lambda_j,\lambda_l}=\frac{1}{2}{\rm tr}\left(\rho\{L_{\lambda_j},L_{\lambda_l}\}\right)=\sum_{k}\sum_{n_k}\frac{(\partial_{\lambda_j}\rho_{\beta,\mu}^{(n_k)})(\partial_{\lambda_l}\rho_{\beta,\mu}^{(n_k)})}{\rho_{\beta,\mu}^{(n_k)}},
\end{equation}
namely

\begin{eqnarray}
\label{entries1} & & F_{\mu,\mu}=\beta^2\Delta^2 N=\beta^2\sum_k\frac{e^{\beta(\varepsilon_k-\mu)}}{(e^{\beta(\varepsilon_k-\mu)}\pm 1)^2}, \\
\label{entries2}& & F_{\beta,\beta}=\Delta^2 (\mu N-H)=\sum_k(\mu-\varepsilon_k)^2\frac{e^{\beta(\varepsilon_k-\mu)}}{(e^{\beta(\varepsilon_k-\mu)}\pm 1)^2}, \\
\label{entries3}& & F_{\mu,\beta}=F_{\beta,\mu}=\beta \, {\rm Cov}(N,\mu N-H)=\beta\sum_k(\mu-\varepsilon_k)\frac{e^{\beta(\varepsilon_k-\mu)}}{(e^{\beta(\varepsilon_k-\mu)}\pm 1)^2},
\end{eqnarray}

\noindent
where the plus signs hold for fermions and the minus signs hold for bosons.

The tensor product in the grand canonical state (\ref{grandcan}), the tensor
product in the symmetric logarithmic derivatives (\ref{sld}), and the sum over
the modes in the entries of the Fisher matrix (\ref{fisher}) witness the
lack of correlations in the mode representation
\cite{Zanardi2001,Narnhofer2004,Barnum2004,Barnum2005,Benatti2012-1,Benatti2012-2}.
The estimation of $(\beta,\mu)$ looks like a classical problem, but the
classical scaling of the Fisher information may no longer hold, because the
number of particles is not fixed. For instance, since the state
(\ref{grandcan}) is separable with respect to the modes, standard arguments
imply that the Fisher information scales linearly with the number of modes,
as the sum in (\ref{fisher}) suggests. However, this scaling does not
correspond to a linear scaling in the average number of particles. Indeed,
the state (\ref{grandcan}) is not in the convex hull of products of
single-particle density matrices, because of the symmetrization or
anti-symmetrization of the Hilbert space. Therefore, the estimation of
$(\beta,\mu)$ with ideal quantum gases can be viewed as an abstract classical
estimation problem with a fluctuating number of particles. We
will show that this estimation is characterized by superlinear scalings of
the Fisher matrix, which cannot result from the same estimation performed
with classical gases.

\section{Bosonic and fermionic gases} \label{sec.gases}

In this section, we present the basic physical quantities of quantum gases,
that will be employed later in the discussion of estimating $(\beta,\mu)$
within several settings. We focus on non-condensed bosonic and fermionic
ideal gases in 
$d$ dimensions spatially confined ($=1,2,3$) either by a box with flat potential and periodic boundary conditions or by a harmonic potential. Bose-Einstein condensation shall be discussed in
section \ref{sec.BEC}.  

\subsection{Homogeneous ideal gases}

Homogeneous ideal gases in $d$ ($=1,2,3$) dimensions are confined in a
parallelepiped shaped box of volume $V_d$, where $V_1=L_x$, $V_2=L_x L_y$,
and $V_3=L_x L_y L_z$ with periodic boundary conditions. The single particle
energies are $\varepsilon_k=(k_x^2+k_y^2+k_z^2)/(2m)$ with
$k_{x,y,z}=\frac{2\pi\hslash}{L_{x,y,z}}n_{x,y,z}$, and $n_{x,y,z}$ running
over all the integers. In the limit of large volume $L_{x,y,z}\to\infty$,
the vector $\vec k$ is approximated by a continuous variable $\vec p$, and
the sum over the modes by an integral 
\begin{equation} \label{cont-hom}
\sum_{\vec k}=\frac{V_d}{(2\pi\hslash)^d}\int_0^\infty d^d\vec p.
\end{equation}
This replacement is exactly the definition of the Riemann integral for
infinity volumes $V_d$, since the energy spacings are vanishingly small, and
is a good approximation at large but finite size. However, the continuum
approximation breaks down when bosonic gases approach the phase transition
to 
Bose-Einstein condensation \cite{Huang,PethickSmith,PitaevkiiStringari} from
high temperatures. 

The average total number of particles and the average
energy are, respectively, 
\begin{eqnarray}
\label{N-hom} \langle N\rangle_d^{\rm hom} & = &
\begin{cases}
\displaystyle -\frac{V_d}{\lambda_T^d} \, {\rm Li}_{\frac{d}{2}}(-e^{\beta\mu}) & \mbox{for femions} \\
\displaystyle \frac{V_d}{\lambda_T^d} \, {\rm Li}_{\frac{d}{2}}(e^{\beta\mu}) & \mbox{for bosons}
\end{cases}, \\
\label{H-hom} \langle H\rangle_d^{\rm hom} & = &
\begin{cases}
\displaystyle -\displaystyle \frac{dV_d}{2\beta\lambda_T^d} \, {\rm Li}_{\frac{d}{2}+1}(-e^{\beta\mu}) & \mbox{for femions} \\
\displaystyle \frac{dV_d}{2\beta\lambda_T^d} \, {\rm Li}_{\frac{d}{2}+1}(e^{\beta\mu}) & \mbox{for bosons}
\end{cases},
\end{eqnarray}

\noindent
where $\lambda_T=\sqrt{2\pi\hslash^2\beta/m}$ is the thermal wavelength and
${\rm Li}_\alpha(z)=\sum_{k=1}^\infty z^k/k^\alpha$ is the polylogarithm \cite{Wood1992}. The entries of the Fisher matrix can
be calculated using (\ref{entries1}-\ref{entries3}), relations (\ref{rel1}-\ref{rel3}), and the property of the polylogarithm $z\frac{\partial {\rm Li}_\alpha (z)}{\partial z}={\rm Li}_{\alpha-1}(z)$. The final results are

\begin{equation}
\label{mu-mu-hom}
(F_d^{\rm hom})_{\mu,\mu}=
\begin{cases}
\displaystyle -\frac{\beta^2 V_d}{\lambda_T^d}{\rm Li}_{\frac{d}{2}-1}(-e^{\beta\mu}) & \mbox{for fermions} \\
\displaystyle \frac{\beta^2 V_d}{\lambda_T^d}{\rm Li}_{\frac{d}{2}-1}(e^{\beta\mu}) & \mbox{for bosons}
\end{cases},
\end{equation}

\begin{equation}
\label{beta-beta-hom}
(F_d^{\rm hom})_{\beta,\beta}=
\begin{cases}
\displaystyle \frac{V_d}{\beta^2\lambda_T^d}\left(-\beta^2\mu^2{\rm Li}_{\frac{d}{2}-1}(-e^{\beta\mu})+d\beta\mu{\rm Li}_{\frac{d}{2}}(-e^{\beta\mu})-\frac{d^2+2d}{4}{\rm Li}_{\frac{d}{2}+1}(-e^{\beta\mu})\right) & \mbox{for fermions} \\
\displaystyle \frac{V_d}{\beta^2\lambda_T^d}\left(\beta^2\mu^2{\rm Li}_{\frac{d}{2}-1}(e^{\beta\mu})-d\beta\mu{\rm Li}_{\frac{d}{2}}(e^{\beta\mu})+\frac{d^2+2d}{4}{\rm Li}_{\frac{d}{2}+1}(e^{\beta\mu})\right) & \mbox{for bosons}
\end{cases},
\end{equation}

\begin{equation}
\label{mu-beta-hom}
(F_d^{\rm hom})_{\mu,\beta}=
\begin{cases}
\displaystyle \frac{V_d}{\lambda_T^d}\left(-\beta\mu{\rm Li}_{\frac{d}{2}-1}(-e^{\beta\mu})+\frac{d}{2}{\rm Li}_{\frac{d}{2}}(-e^{\beta\mu})\right) & \mbox{for fermions} \\
\displaystyle \frac{V_d}{\lambda_T^d}\left(\beta\mu{\rm Li}_{\frac{d}{2}-1}(e^{\beta\mu})-\frac{d}{2}{\rm Li}_{\frac{d}{2}}(e^{\beta\mu})\right) & \mbox{for bosons}
\end{cases}.
\end{equation}

Notice that the two-dimensional case can be a bit simplified, realizing that ${\rm Li}_1(z)=-\ln(1-z)$. For example,
\begin{equation} \label{N-hom2D} 
\langle N\rangle_2^{\rm hom}=
\begin{cases}
\displaystyle \frac{L_x L_y}{\lambda_T^d}\ln(1+e^{\beta\mu}) & \mbox{for femions} \\
\displaystyle -\frac{L_x L_y}{\lambda_T^d}\ln(1-e^{\beta\mu}) & \mbox{for bosons}
\end{cases},
\end{equation}
\noindent
and, for instance, the Fisher information relative to the chemical potential
can be explicitly written as a function of the average number of particles 

\begin{equation} \label{mu-mu-hom2D}
(F_2^{\rm hom})_{\mu,\mu}=
\begin{cases}
\displaystyle \frac{\beta^2 L_x L_y}{\lambda_T^2}\frac{1}{1+e^{-\beta\mu}}=\frac{\beta^2 L_x L_y}{\lambda_T^2}\left(1-e^{-\lambda_T^2\frac{\langle N\rangle_2^{\rm hom}}{L_x L_y}}\right) & \mbox{for fermions} \\
\displaystyle \frac{\beta^2 L_x L_y}{\lambda_T^2}\frac{1}{e^{-\beta\mu}-1}=\frac{\beta^2 L_x L_y}{\lambda_T^2}\left(e^{\lambda_T^2\frac{\langle N\rangle_2^{\rm hom}}{L_x L_y}}-1\right) & \mbox{for bosons}
\end{cases}.
\end{equation}

In the next sections we shall discuss these general formulas in different
regimes. In particular we will elucidate limitations and implications of the
apparent exponential scaling in (\ref{mu-mu-hom2D}). 

\subsection{Harmonically trapped ideal gases}

Now we discuss ideal gases confined in a harmonic potential in $d$
($=1,2,3$) dimensions, with frequencies $\omega_{x,y,z}$ in the three
directions. The single particles energies are
$\varepsilon_k=\hslash(\omega_x n_x+\omega_y n_y+\omega_z n_z)$, with
integers $n_{x,y,z}\geqslant 0$. We define the geometric average of the
frequencies, $\Omega_1=\omega_x$, $\Omega_2=\sqrt{\omega_x \omega_y}$, and
$\Omega_3=(\omega_x \omega_y \omega_z)^{1/3}$. For small frequencies
$\omega_{x,y,z}$, the vector $\vec n$ can be approximated by a continuous
variable $\vec x$ and the sum over the modes becomes an integral 

\begin{equation} \label{cont-harm}
\sum_{\vec n}=\frac{1}{(\hslash\Omega_d)^d}\int_0^\infty d^d\vec x.
\end{equation}

\noindent
As for homogeneous gases, the continuum limit is the exact definition of the
Riemann integral for a vanishing confinement volume $\omega_{x,y,z}\to
0$. It 
is a good approximation at large but finite size, and breaks when
Bose-Einstein condensation occurs
\cite{deGroot1950,Mullin1997,PethickSmith}. Bose-Einstein condensation will
be discussed in the next section. The average total number of particles and
the average energy are respectively 

\begin{eqnarray}
\label{N-harm} \langle N\rangle_d^{\rm harm} & = &
\begin{cases}
\displaystyle -\frac{{\rm Li}_d(-e^{\beta\mu})}{(\beta\hslash\Omega_d)^d} & \mbox{for fermions} \\
\displaystyle \frac{{\rm Li}_d(e^{\beta\mu})}{(\beta\hslash\Omega_d)^d} & \mbox{for bosons}
\end{cases}, \\
\label{H-harm} \langle H\rangle_d^{\rm harm} & = &
\begin{cases}
\displaystyle -\frac{d \, {\rm Li}_{d+1}(-e^{\beta\mu})}{\beta^{d+1}(\hslash\Omega_d)^d} & \mbox{for femions} \\
\displaystyle \frac{d \, {\rm Li}_{d+1}(e^{\beta\mu})}{\beta^{d+1}(\hslash\Omega_d)^d} & \mbox{for bosons}
\end{cases}.
\end{eqnarray}

\noindent
The entries of the Fisher matrix can be calculated similarly to the homogeneous gases, resulting in

\begin{equation}
\label{mu-mu-harm}
(F_d^{\rm harm})_{\mu,\mu}=
\begin{cases}
\displaystyle -\frac{{\rm Li}_{d-1}(-e^{\beta\mu})}{\beta^{d-2}(\hslash\Omega_d)^d} & \mbox{for fermions} \\
\displaystyle \frac{{\rm Li}_{d-1}(e^{\beta\mu})}{\beta^{d-2}(\hslash\Omega_d)^d} & \mbox{for bosons}
\end{cases},
\end{equation}

\begin{equation}
\label{beta-beta-harm}
(F_d^{\rm harm})_{\beta,\beta}=
\begin{cases}
\displaystyle \frac{1}{\beta^{d+2}(\hslash\Omega_d)^d}\left(-\beta^2\mu^2{\rm Li}_{d-1}(-e^{\beta\mu})+2d\beta\mu{\rm Li}_{d}(-e^{\beta\mu})-(d^2+d){\rm Li}_{d+1}(-e^{\beta\mu})\right) & \mbox{for fermions} \\
\displaystyle \frac{1}{\beta^{d+2}(\hslash\Omega_d)^d}\left(\beta^2\mu^2{\rm Li}_{d-1}(e^{\beta\mu})-2d\beta\mu{\rm Li}_{d}(e^{\beta\mu})+(d^2+d){\rm Li}_{d+1}(e^{\beta\mu})\right) & \mbox{for bosons}
\end{cases},
\end{equation}

\begin{equation}
\label{mu-beta-harm}
(F_d^{\rm harm})_{\mu,\beta}=
\begin{cases}
\displaystyle \frac{1}{(\beta\hslash\Omega_d)^d}\left(-\beta\mu{\rm Li}_{d-1}(-e^{\beta\mu})+d{\rm Li}_d(-e^{\beta\mu})\right) & \mbox{for fermions} \\
\displaystyle \frac{1}{(\beta\hslash\Omega_d)^d}\left(\beta\mu{\rm Li}_{d-1}(e^{\beta\mu})-d{\rm Li}_d(e^{\beta\mu})\right) & \mbox{for bosons}
\end{cases}.
\end{equation}

For one-dimensional gases, the average number of particles and the Fisher information $(F_1^{\rm harm})_{\mu,\mu}$ can be written as elementary functions, by means of ${\rm Li}_1(z)=-\ln(1-z)$. Thus, we can explicitly write the dependence of the Fisher matrix on $\langle N\rangle_1^{\rm harm}$: for instance

\begin{equation} \label{N-harm1D}
\langle N\rangle_1^{\rm harm}=
\begin{cases}
\displaystyle \frac{\ln(1+e^{\beta\mu})}{\beta\hslash\omega_x} & \mbox{for fermions} \\
\displaystyle -\frac{\ln(1-e^{\beta\mu})}{\beta\hslash\omega_x} & \mbox{for bosons}
\end{cases},
\end{equation}

\begin{equation} \label{mu-mu-harm1D}
(F_1^{\rm harm})_{\mu,\mu}=
\begin{cases}
\displaystyle \frac{\beta}{\hslash\omega_x}\frac{1}{1+e^{-\beta\mu}}=\frac{\beta}{\hslash\omega_x}\left(1-e^{-\beta\hslash\omega_x\langle N\rangle_1^{\rm harm}}\right) & \mbox{for fermions} \\
\displaystyle \frac{\beta}{\hslash\omega_x}\frac{1}{e^{-\beta\mu}-1}=\frac{\beta}{\hslash\omega_x}\left(e^{\beta\hslash\omega_x\langle N\rangle_1^{\rm harm}}-1\right) & \mbox{for bosons}
\end{cases}.
\end{equation}

As for homogeneous gases, the general formulas for the Fisher matrix shall
be discussed in the next sections within different physical regimes. 

\section{Fisher matrix in the continuum approximation} \label{sec.cont}

In this section we describe the sensitivity of the estimation of
$(\beta,\mu)$ for quantum gases. In statistical mechanics, the thermodynamic
limit is usually considered, meaning an infinite number of particles and
an infinite confinement volume such that the density is fixed. We shall
discuss how the Fisher matrix scales with the average number of particles,
approaching the thermodynamical limit. 

A different assumption is to fix the confinement volume, rather than the
density, which is natural in mesoscopic systems, such as in experiments with
atomic gases, as reported in
\cite{Gorlitz2001,Greiner2001,Esteve2006,vanAmerongen2008,vanAmerongen2008-2,Bouchoule2011,Armijo2011}.
In particular, when a  gas is confined in one or two dimensions with strong
confinements in the remaining directions, the number of particles is limited
\cite{Gorlitz2001} but still large ($\sim 10^4,10^5$) for quantum
metrological applications. Moreover, the above mentioned experiments were
performed with finite confinement volumes which provide high particle
densities. In this framework, all the results are formally the same as in
the thermodynamic limit. The only difference is that the density appears
not just as a prefactor but enters in the scalings, as it is proportional to
the number of particles. 

\subsection{Homogeneous ideal gases}

The thermodynamic limit of homogeneous gases is defined as $\langle
N\rangle_d^{\rm hom}\to\infty$ and $L_{x,y,z}\to\infty$ with
$\varrho=\langle N\rangle_d^{\rm hom}/V_d$ finite. All the quantities
(\ref{N-hom},\ref{H-hom},\ref{mu-mu-hom},\ref{beta-beta-hom},\ref{mu-beta-hom}) are
linear in the volume. Therefore, all the entries of the Fisher matrix scale
linearly with $\langle N\rangle_d^{\rm hom}$. In order to discuss the
prefactors $(F_d^{\rm hom})_{\#,\#}/\langle N\rangle$, we recall that there
are no restrictions for $\mu\in[-\infty,\infty]$ and $\beta\in[0,\infty]$
for fermions, while  for bosons in the non-condensed phase
$\mu\in[-\infty,0)$ and $\beta$ is larger than the 
  critical inverse temperature. The
  polylogarithms involved are bounded for all finite values of their
  argument, except ${\rm Li}_\alpha(z)$ with ${\rm Re}(\alpha)<1$ that
  diverges for $z=1$. Thus, the prefactors $(F_d^{\rm hom})_{\#,\#}/\langle
  N\rangle_d^{\rm hom}$ are finite almost always but for some exceptional
  points that we are going to discuss. 

\paragraph{Classical limit}

First we show that in the classical limit, i.e. high temperature and low
density implying $e^{\beta\mu}\ll 1$, the shot-noise regime typical of
classical statistics is recovered. With the help of the asymptotics ${\rm
  Li}_{\alpha}(z)\simeq z$ for all $\alpha$ and $|z|\ll 1$, we easily
compute statistical averages and the Fisher matrix in the classical limit: 

\begin{eqnarray}
\label{class.hom.1} && \langle N\rangle_d^{\rm hom}\simeq\frac{V_d}{\lambda_T^d}e^{\beta\mu}, \qquad \langle H\rangle_d^{\rm hom}\simeq\frac{d V_d}{2\beta\lambda_T^d}e^{\beta\mu}, \qquad (F_d^{\rm hom})_{\mu,\mu}\simeq\beta^2\langle N\rangle_d^{\rm hom}, \\
\label{class.hom.2} && (F_d^{\rm hom})_{\beta,\beta}\simeq\langle N\rangle_d^{\rm hom}\left(\mu^2-d\frac{\mu}{\beta}+\frac{d^2+2d}{4\beta^2}\right), \qquad (F_d^{\rm hom})_{\mu,\beta}\simeq\langle N\rangle_d^{\rm hom}\left(\beta\mu-\frac{d}{2}\right).
\end{eqnarray}

\noindent
The entries of the Fisher matrix scale linearly with the average number of
particles, and are exactly the same as those of homogeneous classical gases
derived in appendix \ref{class.lim}. 

The polylogarithms in the Fisher matrix continuously and monotonically
increase when $e^{\beta\mu}$ increases, i.e. going away from the classical
limit. If the polylogarithms are bounded, the corresponding change in the
Fisher matrix, and thus in the sensitivity of the measurements, is only a
numerical prefactor that does not modify the scaling with the average number
of particles. In the following, we discuss limits where the Fisher matrix
deviates from the shot-noise typical of the classical limit.  

\paragraph{Low temperatures: fermionic gases}

\begin{figure}[htbp]
\centering
\includegraphics[width=0.4\columnwidth]{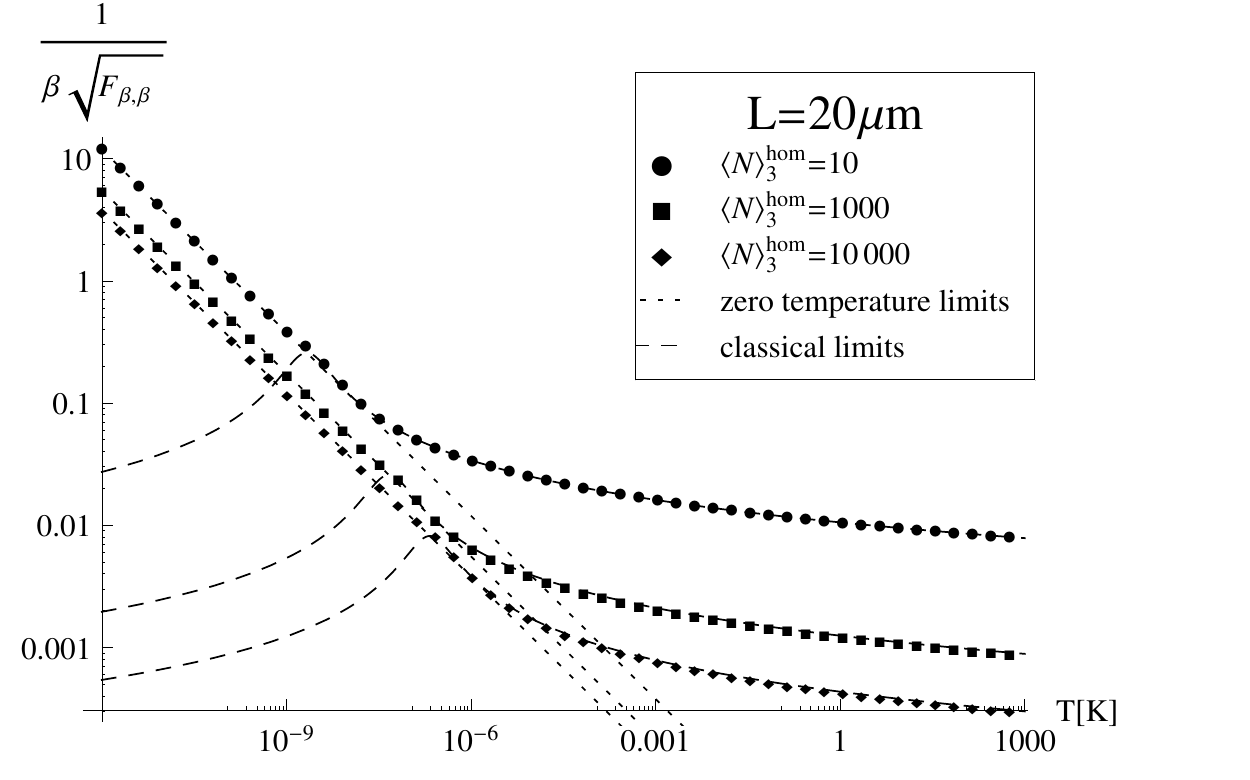}
\includegraphics[width=0.4\columnwidth]{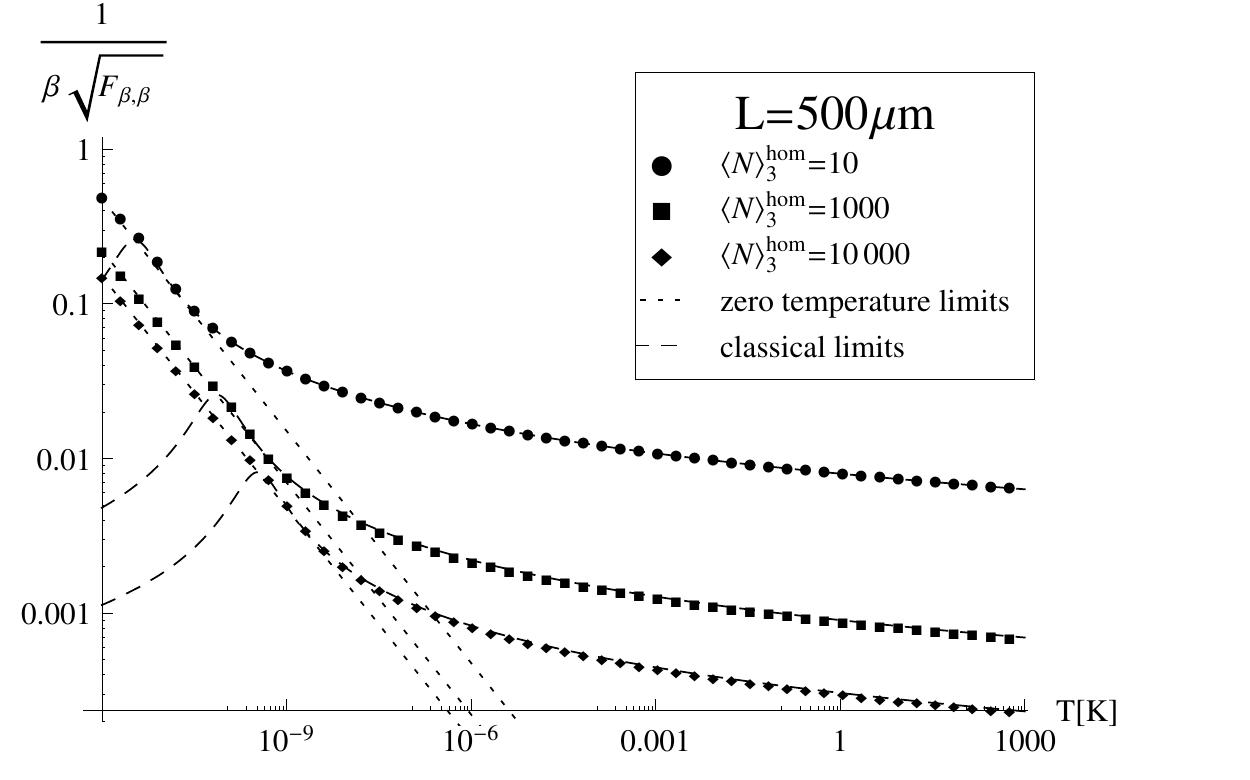}
\caption{Log-log-plot of the optimal relative error $1/(\beta\sqrt{(F_d^{\rm hom})_{\beta,\beta}})$ for a homogenous gas of $^6$Li atoms in three dimensions as function of temperature: $L=20\mu$m (left) and $L=500\mu$m  (right); and $\langle N\rangle_3^{\rm hom}=10$ (circles), $\langle N\rangle_3^{\rm hom}=1000$ (squares), and $\langle N\rangle_3^{\rm hom}=10000$ (diamonds). Dotted (dashed) lines are the corresponding small temperature limits \eqref{zero.temp.homf} (classical limits (\ref{class.hom.1},\ref{class.hom.2})). For a given temperature, we computed the chemical potential inverting the equation of state \eqref{N-hom} with the Netown's method, then we used this result in \eqref{beta-beta-hom} and plotted the corresponding values of $1/(\beta\sqrt{(F_d^{\rm hom})_{\beta,\beta}})$.
}
\label{temp}
\end{figure}

\begin{figure}[htbp]
\centering
\includegraphics[width=0.4\columnwidth]{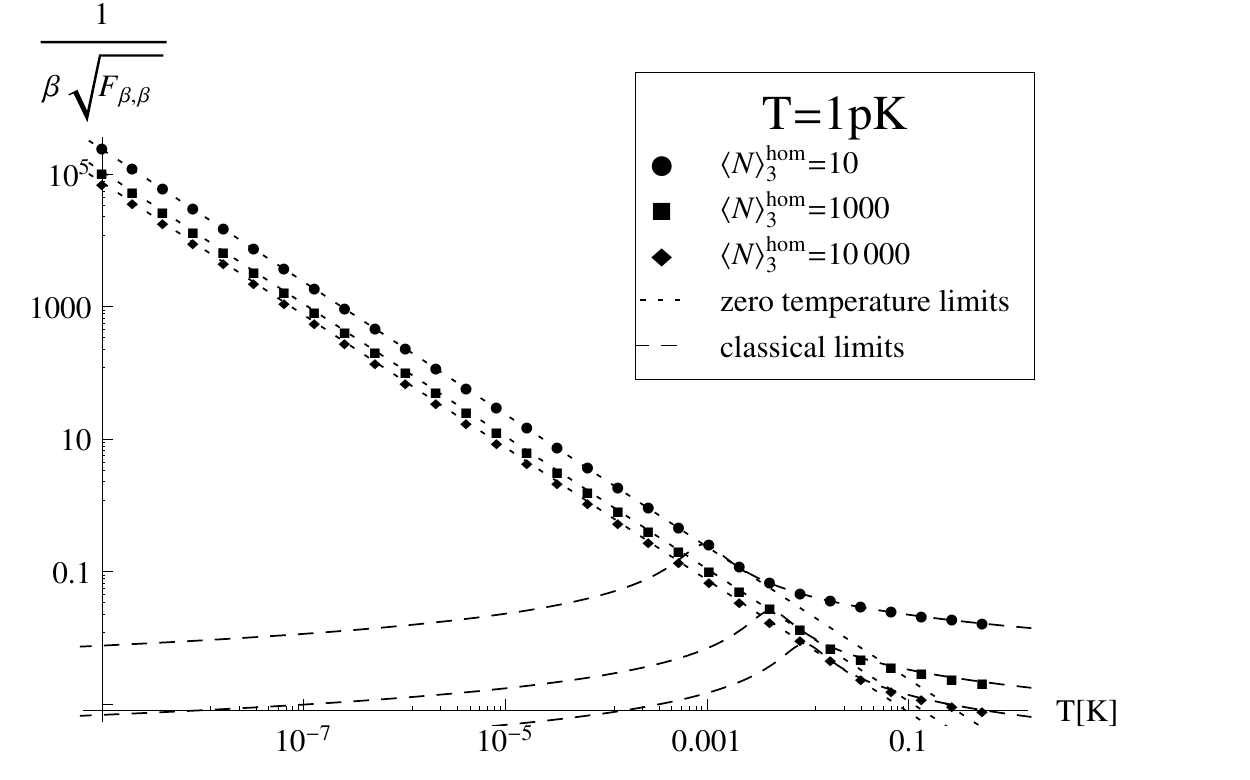}
\includegraphics[width=0.4\columnwidth]{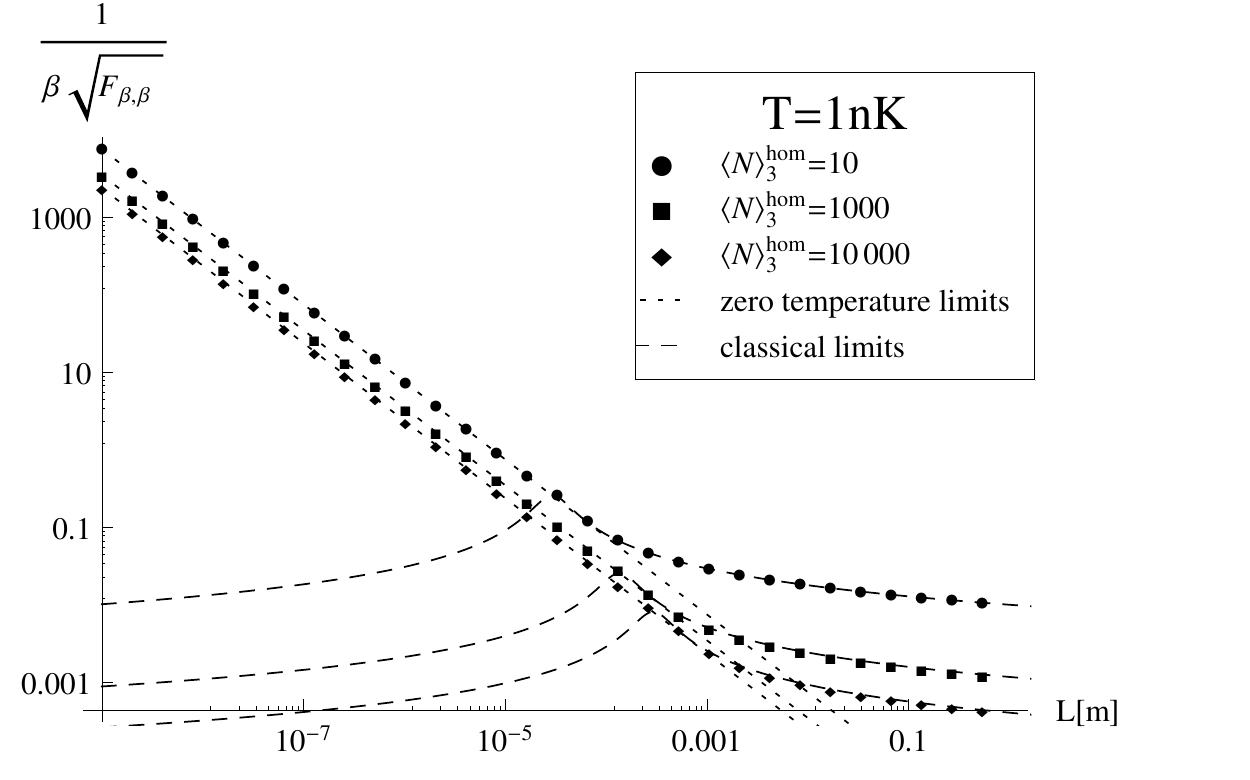}
\includegraphics[width=0.4\columnwidth]{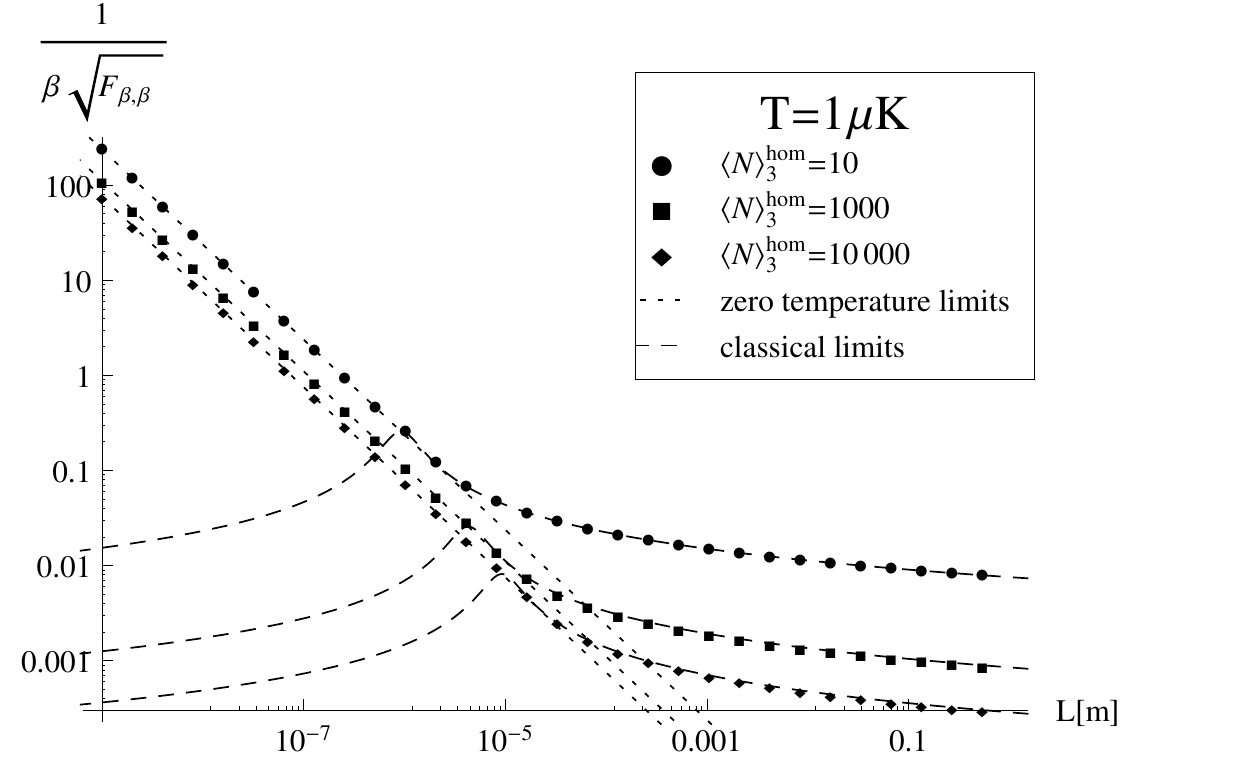}
\includegraphics[width=0.4\columnwidth]{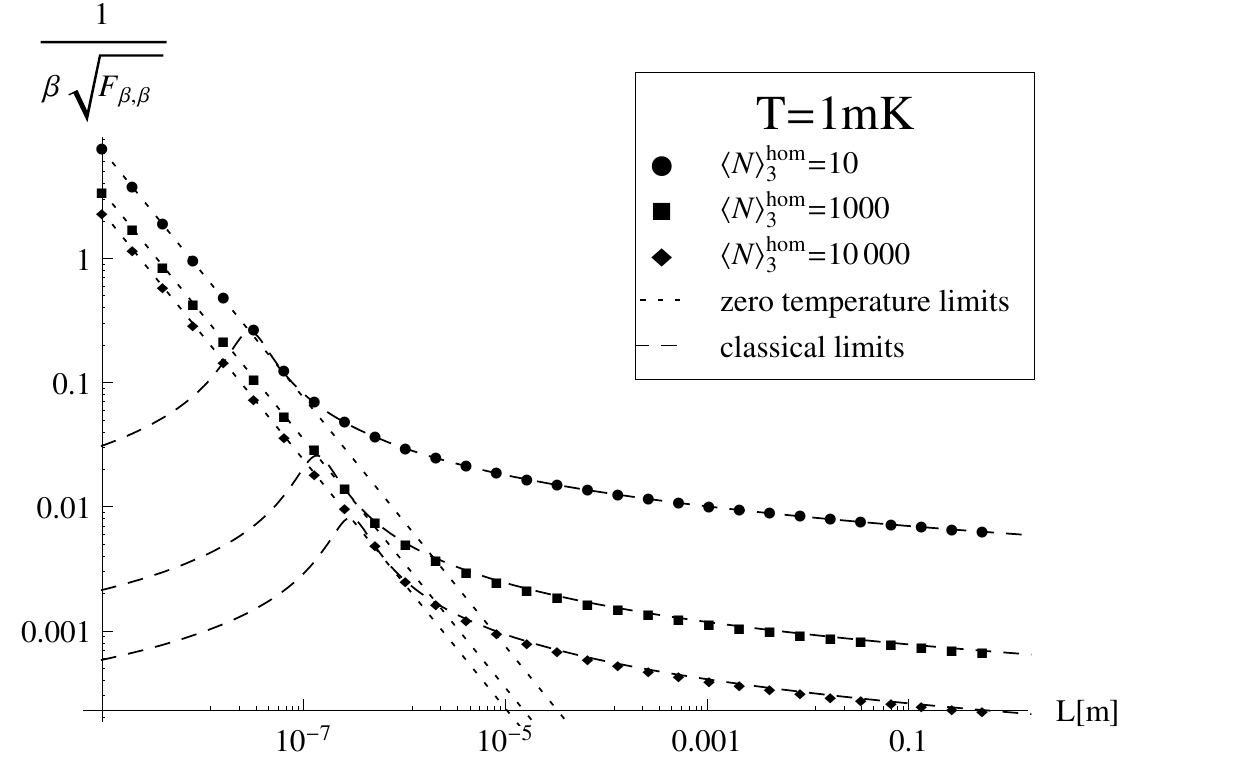}
\caption{Log-log plot of the optimal relative error $1/(\beta\sqrt{(F_d^{\rm hom})_{\beta,\beta}})$ for a homogenous gas of $^6$Li atoms in three dimensions: $T=1$pK (up, left), $T=1$nK (up, right), $T=1\mu$K (down, left), and $T=1$mK  (down, right); and $\langle N\rangle_3^{\rm hom}=10$ (circles), $\langle N\rangle_3^{\rm hom}=1000$ (squares), and $\langle N\rangle_3^{\rm hom}=10000$ (diamonds). Dotted (dashed) lines are the corresponding small temperature limits \eqref{zero.temp.homf} (classical limits (\ref{class.hom.1},\ref{class.hom.2})). For a given $L$, we computed the chemical potential inverting the equation of state \eqref{N-hom} with the Netown's method, then we used this result in \eqref{beta-beta-hom} and plotted the corresponding values of $1/(\beta\sqrt{(F_d^{\rm hom})_{\beta,\beta}})$.  
}
\label{size}
\end{figure}

The small temperature limit of fermionic gases, $\beta\to\infty$ and $\mu\to
E_F$ where $E_F$ is the Fermi energy, together with the property ${\rm
  Li}_{\alpha}(-e^x)=-x^\alpha/\Gamma(\alpha+1)-\pi^2
x^{\alpha-2}/(6\Gamma[\alpha-1])+{\cal O}(x^{\alpha-4})$ for $Re(x)\gg 1$
\cite{Wood1992}, provides the following Fisher matrix:

\begin{equation} \label{zero.temp.homf}
(F_d^{\rm hom})_{\mu,\mu}\simeq\frac{d\beta}{2\mu}\langle N\rangle_d^{\rm hom}, \qquad (F_d^{\rm hom})_{\beta,\beta}\simeq\frac{d\pi^2}{6\beta^3\mu}\langle N\rangle_d^{\rm hom}, \qquad (F_d^{\rm hom})_{\mu,\beta}\simeq\frac{(2-d)d\pi^2}{12\beta^2\mu^2}\langle N\rangle_d^{\rm
  hom}.
\end{equation}
Hence, the temperature can be measured only with a very bad sensitivity which is bounded by the inverse of the Fisher information $(F_d^{\rm hom})_{\beta,\beta}$, according to the Cram\'er-Rao bound \eqref{CRB}. The Fisher information per particle $(F_d^{\rm
  hom})_{\beta,\beta}/\langle N\rangle_d^{\rm hom}$ vanishes as $T^3$
and the relative error of the optimal estimation
$1/(\beta\sqrt{(F_d^{\rm hom})_{\beta,\beta}})\leqslant\sqrt{\rm
  var(\beta)}/\beta=\Delta T/T$ diverges as $1/\sqrt{T}$ for $T\to
0$. In spite of this limit for arbitrary small temperatures, the best
relative error $1/(\beta\sqrt{(F_d^{\rm hom})_{\beta,\beta}})$ is
small for experimentally relevant settings, as shown as a function of
the temperature and the size in figures \ref{temp} and \ref{size} for
three-dimensional homogeneous gases of $^6$Li atoms. These values fit
the small temperature \eqref{zero.temp.homf} (classical
(\ref{class.hom.1},\ref{class.hom.2})) scaling for large (small)
temperatures and sizes, and bound the relative errors of actual
thermometry experiments: the optimal relative errors are one order of
magnitude smaller than those obtained in experiments with similar
physical settings \cite{Muller2010,Sanner2010}. It is noticeable that
the best relative error in the classical limit is much smaller than
that in the quantum regime for very small temperatures. However,
quantum effects at those temperature cannot be neglected, and the
consideration of classical gases is just an academic problem. 

On the other hand, the chemical
potential can be 
measured with almost no error due to the divergence of $(F_d^{\rm
  hom})_{\mu,\mu}/\langle N\rangle_d^{\rm hom}$ as $1/T$ for small $T$. If
the temperature is 
exactly zero, all the 
fermions are frozen in the lowest energies up to the Fermi energy. The
smallest change in the chemical potential equals the spacing between the
Fermi energy and the next excited state, keeping the temperature fixed. The
state consequently changes into an orthogonal state with a different number
of particles. This non-smooth change is in contrast with the assumptions upon which the standard quantum Cram\'er-Rao bound \eqref{CRB} is based, and requires the generalization
to 
non-differentiable models \cite{Tsuda2005}. However, this gives an intuition
for the divergence of $(F_d^{\rm hom})_{\mu,\mu}/\langle N\rangle_d^{\rm
  hom}$ in the regime of small temperatures. In this context, remember that
temperature and chemical potential can be jointly measured. We discuss the zero temperature case in more detail in appendix \ref{0temp}.

\paragraph{Low temperatures: bosonic gases}

In the low-temperature bosonic case (but above the condensation
  temperature), some of the polylogarithms diverge as ${\rm 
  Li}_{\alpha}(e^x)=\Gamma(1-\alpha)(-x)^{\alpha-1}+{\cal O}(1)$ for $x\to
0$ and ${\rm Re}(\alpha)<1$. Thus, the prefactors $(F_d^{\rm
  hom})_{\beta,\beta}/\langle N\rangle_d^{\rm hom}$ and $(F_d^{\rm
  hom})_{\mu,\beta}/\langle N\rangle_d^{\rm hom}$ are finite or zero for all
values of the fugacity $e^{\beta\mu}$ in $[0,1]$, while $(F_d^{\rm
  hom})_{\mu,\mu}/\langle N\rangle_d^{\rm hom}$ diverges when the fugacity
approaches one, namely its value at the critical temperature
\cite{Huang,PethickSmith,PitaevkiiStringari}, implying a very precise
measurement of the chemical potential. These divergences are ruled by the
way the chemical potential approaches zero. In this limit, we derive
upper bounds for the Fisher information $(F_d^{\rm hom})_{\mu,\mu}$ and the
corresponding scaling with $\langle N\rangle_d^{\rm hom}$. Since the
polylogarithms are continuous functions, the Fisher information $(F_d^{\rm
  hom})_{\mu,\mu}$ assumes all values between (\ref{class.hom.1}) in the
classical limit and the following
bounds (up to inequality (\ref{mu-hom-1D-2}))
which are saturated
 at the onset of the BEC.

The value of the chemical potential is constrained by two conditions. The
first one is implied by the application of the continuum approximation
(\ref{cont-hom}). Indeed, approximating the sum with the integral is valid
only for very small momentum spacings, namely small
$2\pi\hslash/L_{x,y,z}$, which is the  quantity that becomes infinitesimal in the continuum limit. Since $\beta\mu$  is always subtracted
from $\beta\varepsilon_k$, the validity of the continuum
approximation requires that  $-\beta\mu$ cannot be smaller than the energy
spacing, i.e. 

\begin{equation} \label{val-ca-hom}
-\beta\mu\geqslant\frac{\beta(2\pi\hslash)^2}{2mV_d^\frac{2}{d}}=\frac{\lambda_T^2\varrho^\frac{2}{d}}{(\langle N\rangle_d^{\rm hom})^\frac{2}{d}},
\end{equation}

\noindent
for homogeneous gases. If at low temperatures (\ref{val-ca-hom}) is violated
one should set $-\beta\mu=0$ in all thermodynamical quantities which do not
diverge under this replacement in the continuum approximation. For thermodynamical quantities that would diverge when
setting $-\beta\mu=0$ one has to estimate
the exact sums over the modes without the
continuum approximation, in order to carefully estimate their scaling with
the average number of particles.

A second bound is given by the occupation of the
ground state: 

\begin{equation} \label{num-const-hom}
\langle a_0^\dag a_0\rangle=\frac{1}{e^{-\beta\mu}-1}
\leqslant\langle N\rangle_d^{\rm hom}\Rightarrow -\beta\mu\geqslant\ln\left(1+\frac{1}{\langle N\rangle_d^{\rm hom}}\right)\simeq\frac{1}{\langle N\rangle_d^{\rm hom}}.
\end{equation}

In three dimensions, the continuum approximation through (\ref{val-ca-hom}) bounds the Fisher information to

\begin{equation} \label{mu-hom-3D}
(F_3^{\rm hom})_{\mu,\mu}\lesssim\frac{\beta^2}{\lambda_T^4\varrho^\frac{4}{3}}(\langle N\rangle_3^{\rm hom})^\frac{4}{3}\simeq\frac{\beta^2}{\zeta\left(\frac{3}{2}\right)^\frac{4}{3}}(\langle N\rangle_3^{\rm hom})^\frac{4}{3},
\end{equation}

\noindent
where the last inequality follows from $\lambda_T^3\rho\simeq\zeta(3/2)$ for small chemical potentials. Smaller chemical potentials, $\ln(1+1/\langle N\rangle_3^{\rm hom})\leqslant-\beta\mu\leqslant o(\langle N\rangle_3^{\rm hom})^{-2/3}$, are practically zero in the continuum approximation and $(F_3^{\rm hom})_{\mu,\mu}$ diverges. From the computation of the discrete sums in this regime, the leading contribution yields

\begin{equation} \label{mu-hom-3D-2}
(F_3^{\rm hom})_{\mu,\mu}\simeq\frac{1}{\mu^2}+\frac{\beta^2 V_3^{\frac{4}{3}}}{\pi^2\lambda_T^4}\sum_{(n_x,n_y,n_z)\neq(0,0,0)}\frac{1}{(n_x^2+n_y^2+n_z^2)^2}\simeq\frac{1}{\mu^2}+\frac{16.5\beta^2}{\pi^2\lambda_T^4\varrho^\frac{4}{3}}(\langle N\rangle_3^{\rm hom})^\frac{4}{3}\simeq\frac{1}{\mu^2}+\frac{16.5 \, \beta^2}{\pi^2\zeta\left(\frac{3}{2}\right)^\frac{4}{3}}(\langle N\rangle_3^{\rm hom})^\frac{4}{3}
\end{equation}
for isotropic gases \cite{Giorgini1998,PitaevkiiStringari}, in agreement
with the limiting scaling in the continuum approximation, and where the
constant $16.5$ was estimated from numerical summation ranging from -256 to
256. The last inequality follows from $\lambda_T^3\rho\simeq\zeta(3/2)$ for
small chemical potentials. 

In one and two dimensions, the finiteness of the density $\varrho$ implies
the finiteness of the chemical potential $\mu\neq 0$, and thus the
finiteness of the prefactors $(F_{1,2}^{\rm hom})_{\mu,\mu}/\langle
N\rangle_{1,2}^{\rm hom}$. However, $(F_{1,2}^{\rm hom})_{\mu,\mu}$ exhibits
a superlinear scaling with the average number of particles if the volume is
fixed rather than the density; similarly, allowing large densities $\varrho$
gives large prefactors $(F_{1,2}^{\rm hom})_{\mu,\mu}/\langle
N\rangle_{1,2}^{\rm hom}$. For the two-dimensional case, the explicit
dependence of this prefactor is given by
(\ref{N-hom2D},\ref{mu-mu-hom2D}). The conditions (\ref{val-ca-hom}) and
(\ref{num-const-hom}) give the same scaling for the chemical potential,
which constraints the density, using the estimation of (\ref{N-hom2D}) for
small chemical potentials: $\lambda_T^2\varrho\lesssim\ln\langle
N\rangle_2^{\rm hom}$. The resulting Fisher information is 

\begin{equation} \label{mu-hom-2D}
(F_2^{\rm hom})_{\mu,\mu}\simeq\beta^2\langle N\rangle_2^{\rm hom}\frac{e^{\lambda_T^2\varrho}}{\lambda_T^2\varrho}\lesssim\beta^2\frac{(\langle N\rangle_2^{\rm hom})^2}{\log\langle N\rangle_2^{\rm hom}}.
\end{equation}

\noindent
For extremely high densities in $d=2$ ($\lambda_T^2\varrho>1$), the condition (\ref{val-ca-hom}) is more stringent than (\ref{num-const-hom}). Both the average number of particles and $(F_2^{\rm hom})_{\mu,\mu}$ diverge at zero chemical potentials. For isotropic gases with intermediate chemical potentials between (\ref{val-ca-hom}) and (\ref{num-const-hom}), the dominant contributions of the discrete sums are

\begin{eqnarray}
\langle N\rangle_2^{\rm hom} & \simeq &
-\frac{1}{\beta\mu}+\frac{V_2}{\pi\lambda_T^2}\sum_{\substack{n_x,n_y=0\\
    (n_x,n_y)\neq(0,0)}}^{{\cal
    O}\left(\frac{V_2}{\pi\lambda_T^2}\right)}\frac{1}{n_x^2+n_y^2}\simeq-\frac{1}{\beta\mu}+\frac{V_2}{2\lambda_T^2}\ln\left(\frac{V_2}{\pi\lambda_T^2}\right), \\ 
\label{mu-hom-2D-2} (F_2^{\rm hom})_{\mu,\mu} & \simeq &
\frac{1}{\mu^2}+\frac{\beta^2 V_2^2}{\pi^2\lambda_T^4}\sum_{(n_x,n_y)\neq(0,0)}\frac{1}{(n_x^2+n_y^2)^2}\simeq\frac{1}{\mu^2}+\frac{6.03 \, \beta^2}{\pi^2\lambda_T^4\varrho^2}(\langle N\rangle_2^{\rm hom})^2\lesssim\frac{1}{\mu^2}+\frac{6.03 \, \beta^2}{\pi^2}\frac{(\langle N\rangle_2^{\rm hom})^2}{\ln^2\langle N\rangle_2^{\rm hom}},
\end{eqnarray}

\noindent
where the last inequality holds under the condition $\lambda_T^2\varrho\gtrsim\ln\langle N\rangle_2^{\rm hom}$, opposite to that in the continuum approximation.

In one dimension and at not too large densities $\lambda_T\varrho\lesssim\sqrt{\langle N\rangle_1^{\rm hom}}$, the continuum approximation gives a more stringent constraint (\ref{val-ca-hom}) than the physical requirement (\ref{num-const-hom}). In the continuum approximation, the limit of large density and $\mu$ close to zero, together with the above bound of the density, give
\begin{equation} \label{mu-hom-1D}
(F_1^{\rm hom})_{\mu,\mu}\simeq\frac{\beta^2}{2\pi}\lambda_T^2\varrho^2\langle N\rangle_1^{\rm hom}\lesssim\frac{\beta^2}{2\pi}(\langle N\rangle_1^{\rm hom})^2.
\end{equation}
For higher densities $\lambda_T\varrho\gtrsim\sqrt{\langle N\rangle_1^{\rm hom}}$, there is an intermediate regime between (\ref{val-ca-hom}) and (\ref{num-const-hom}), where the chemical potential should be set to zero for a consistent application of the continuum approximation. However, since both the average number of particles and $(F_1^{\rm hom})_{\mu,\mu}$ diverge at zero chemical potentials, they are estimated by the dominant contributions of the respective discrete sums:
\begin{eqnarray}
\langle N\rangle_1^{\rm hom} & \simeq & -\frac{1}{\beta\mu}+\frac{L_x^2}{\pi\lambda_T^2}\sum_{n_x=1}^{\infty}\frac{1}{n_x^2}=-\frac{1}{\beta\mu}+\frac{\pi L_x^2}{6\lambda_T^2}, \\ 
\label{mu-hom-1D-2} (F_1^{\rm hom})_{\mu,\mu} & \simeq &
\frac{1}{\mu^2}+\frac{\beta^2 L_x^4}{\pi^2\lambda_T^4}\sum_{n_x=1}^{\infty}\frac{1}{n_x^4}=\frac{1}{\mu^2}+\frac{\pi^2\beta^2}{90\lambda_T^4\rho^4}(\langle N\rangle_1^{\rm hom})^4\lesssim\frac{1}{\mu^2}+\frac{\pi^2\beta^2}{90}(\langle N\rangle_1^{\rm hom})^2,
\end{eqnarray}
where the last inequality comes from the above high density condition $\lambda_T\varrho\gtrsim\sqrt{\langle N\rangle_1^{\rm hom}}$.

\begin{figure}[htbp]
\centering
\includegraphics[width=0.8\columnwidth]{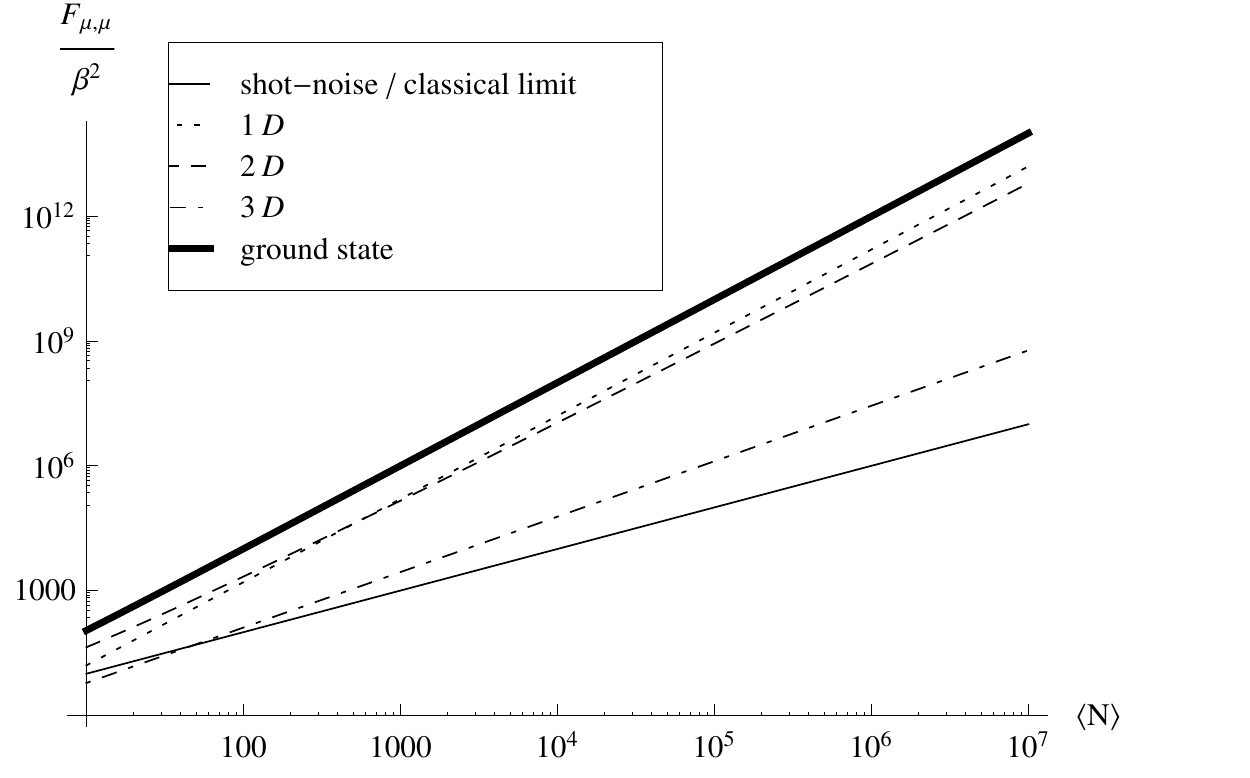}
\caption{Log-log plot of the upper bounds of the rescaled Fisher information
  $(F_d^{\rm hom})_{\mu,\mu}/\beta^2$ for homogenous Bose gas above the
  condensation temperature against the average number of
  particles within the continuum approximation in three dimensions
  (dotdashed line) (\ref{mu-hom-3D}), two dimensions (dashed line)
  (\ref{mu-hom-2D}), and one dimension (dotted line) (\ref{mu-hom-1D}). The
  continuous line is the shot-noise reproduced by the classical gas,
  eq.(\ref{class.hom.1}). The 
  thick line is the Fisher information if all particles are in the ground
  state, namely the zero temperature case, eq.(\ref{eq:T0}).
}
\label{hom_gas}
\end{figure}

In Fig. \ref{hom_gas}, we plot the upper bounds
(\ref{mu-hom-3D},\ref{mu-hom-2D},\ref{mu-hom-1D}) of the Fisher information
$(F_d^{\rm hom})_{\mu,\mu}$ within the continuum approximation versus the
average number of particles in double logarithmic scale. Since the slope
represents the exponent of the dependence of $(F_d^{\rm hom})_{\mu,\mu}$ on
$\langle N\rangle_d^{\rm hom}$, we notice that decreasing the
dimensionality, the sensitivity of the chemical potential increases. We
compare the curves with the linear scaling (shot-noise) reproduced by the
classical limit and the zero temperature case where all the particles occupy
the ground state (see section \ref{sec.BEC}).

\subsection{Harmonically trapped ideal gases}

For gases confined in a harmonic potential, the thermodynamic limit is
defined as $\langle N\rangle_d^{\rm harm}\to\infty$ and $\omega_{x,y,z}\to
0$. In order to have a finite density either the quantity $\langle
N\rangle_d^{\rm harm}\Omega_d^d$ \cite{Mullin1997,Mullin2012} or $\langle
N\rangle_d^{\rm harm}\Omega_d^{\frac{d}{2}}$ \cite{Beau2010} is fixed. This
choice affects the thermodynamics of the gas, such as the phase transition
towards a Bose-Einstein condensate \cite{Beau2010,Mullin2012}. We 
follow the physical arguments presented in \cite{Mullin1997,Mullin2012} and
define
$\tilde\varrho\equiv\langle N\rangle_d^{\rm harm}\Omega_d^d$. Since all the
quantities
(\ref{N-harm},\ref{H-harm},\ref{mu-mu-harm},\ref{beta-beta-harm},\ref{mu-beta-harm}) are
proportional to $\Omega_d^{-d}$, the entries of the Fisher matrix scale
linearly with $\langle N\rangle_d^{\rm harm}$. As before,
$\mu\in[-\infty,\infty]$ and $\beta\in[0,\infty]$ for fermions, while
$\mu\in[-\infty,0)$ and $\beta$ is larger than the critical inverse
  temperature for bosons in the non-condensed phase. With the above
  mentioned properties of the polylogarithms, we can study when the
  prefactors $(F_d^{\rm harm})_{\#,\#}/\langle N\rangle_d^{\rm harm}$
  diverge. 

\paragraph{Classical limit}

In the classical limit, i.e. $e^{\beta\mu}\ll 1$, the statistical averages and the Fisher matrix are

\begin{eqnarray}
\label{class.harm.1} && \langle N\rangle_d^{\rm harm}\simeq\frac{e^{\beta\mu}}{(\beta\hslash\Omega_d)^d}, \qquad \langle H\rangle_d^{\rm harm}\simeq\frac{d \, e^{\beta\mu}}{\beta^{d+1}(\hslash\Omega_d)^d}, \qquad (F_d^{\rm harm})_{\mu,\mu}\simeq\beta^2\langle N\rangle_d^{\rm harm}, \\
\label{class.harm.2} && (F_d^{\rm harm})_{\beta,\beta}\simeq\langle N\rangle_d^{\rm hom}\left(\mu^2-2d\frac{\mu}{\beta}+\frac{d^2+2d}{\beta^2}\right), \qquad (F_d^{\rm harm})_{\mu,\beta}\simeq\langle N\rangle_d^{\rm harm}(\beta\mu-d).
\end{eqnarray}

\noindent
The entries of the Fisher matrix scale linearly with the average number of particles, and equal those of classical gases derived in the appendix \ref{class.lim}.

As for the homogenous gases, we now investigate physical regimes where the Fisher matrix overcomes the linear scaling with $\langle N\rangle_d^{\rm harm}$, which characterizes the classical limit. 

\paragraph{Low temperatures: fermionic gases}

At small temperatures $\beta\to\infty$, the chemical potential of fermionic
gases is close to the Fermi energy $\mu\to E_F$, and

\begin{equation}
(F_d^{\rm harm})_{\mu,\mu}\simeq\frac{d\beta}{\mu}\langle N\rangle_d^{\rm harm}, \qquad (F_d^{\rm harm})_{\beta,\beta}\simeq\frac{d\pi^2}{3\beta^3\mu}\langle N\rangle_d^{\rm harm}, \qquad (F_d^{\rm harm})_{\mu,\beta}\simeq\frac{(1-d)d\pi^2}{3\beta^2\mu^2}\langle N\rangle_d^{\rm harm}.
\end{equation}
As for homogeneous gases,
the sensitivity of a temperature measurement is very bad, whereas the
chemical potential can be measured with infinitely high sensitivity. The
interpretation of the divergence of 
$(F_d^{\rm harm})_{\mu,\mu}$ is the same as for homogeneous fermionic gases:
at zero temperature, all the fermions are frozen in the lowest energies up
to the Fermi energy, the smallest change of the chemical potential is the
spacing between the Fermi energy and the next excited state, and the state
suddenly changes into an orthogonal state with a different number of
particles. Quantum estimation theory for non-differentiable models
\cite{Tsuda2005} needs to be applied when the temperature is exactly zero, as
discussed in the appendix \ref{0temp}. 

\paragraph{Low temperatures: bosonic gases}
For bosonic gases at low temperature, (but above the condensation 
  temperature), the prefactors $(F_d^{\rm harm})_{\beta,\beta}/\langle
  N\rangle_d^{\rm harm}$ and $(F_d^{\rm harm})_{\mu,\beta}/\langle
  N\rangle_d^{\rm harm}$ are always finite or zero. However, $(F_d^{\rm
  harm})_{\mu,\mu}/\langle N\rangle_d^{\rm harm}$ can diverge when the
  temperature approaches the critical temperature
  \cite{deGroot1950,Mullin1997,PethickSmith}, corresponding to $\mu\to 0$. 
This
  stems from the divergence of the polylogarithms already mentioned for
  bosonic homogeneous gases. Also for harmonic gases, we derive upper bounds
  of the scaling of $(F_d^{\rm harm})_{\mu,\mu}/\langle N\rangle_d^{\rm
    harm}$ with respect to $\langle N\rangle_d^{\rm harm}$, which are
  saturated at the onset of BEC. The Fisher information $(F_d^{\rm
    harm})_{\mu,\mu}$ varies continuously 
  between its classical limit (\ref{class.harm.1}) and the following bounds,
  due to the continuity of the polylogarithms. 

As for the homogeneous gases, the chemical potential is bounded by the energy spacing in the continuum approximation, i.e. 

\begin{equation} \label{val-ca-harm}
-\beta\mu\geqslant\beta\hslash\Omega_d=\beta\hslash\left(\frac{\tilde\varrho}{\langle N\rangle_d^{\rm harm}}\right)^{1/d},
\end{equation}

\noindent
for isotropic confinements. The other bound is given by the occupation of the ground state, i.e.

\begin{equation} \label{num-const-harm}
\langle a_0^\dag a_0\rangle=\frac{1}{e^{-\beta\mu}-1}\leqslant\langle N\rangle_d^{\rm harm}\Rightarrow -\beta\mu\geqslant\ln\left(1+\frac{1}{\langle N\rangle_d^{\rm harm}}\right)\simeq\frac{1}{\langle N\rangle_d^{\rm harm}}.
\end{equation}

\noindent
Therefore, for $\ln(1+1/\langle N\rangle_2^{\rm harm})\leqslant-\beta\mu\leqslant o(\langle N\rangle_2^{\rm harm})^{-1/d}$, the continuum approximation can break and the discrete sums must be computed, as discussed for the homogeneous gases.

In three dimensions, both $\langle N\rangle_3^{\rm harm}$ and $(F_3^{\rm harm})_{\mu,\mu}$ remain finite, and for small chemical potentials

\begin{equation} \label{mu-harm-3D}
(F_3^{\rm harm})_{\mu,\mu}\simeq\frac{\pi^2\beta^2}{6\zeta(3)}\langle N\rangle_3^{\rm harm}.
\end{equation}

In two dimensions, the Fisher information relative to the chemical potential diverges as $(F_2^{\rm harm})_{\mu,\mu}/\langle N\rangle_2^{\rm harm}\simeq-\frac{6}{\pi^2}\beta^2\ln(-\beta\mu)$. The condition (\ref{val-ca-harm}) from the continuum approximation yields

\begin{equation} \label{mu-harm-2D}
(F_2^{\rm harm})_{\mu,\mu}\lesssim\frac{3\beta^2}{\pi^2}\langle N\rangle_2^{\rm harm}\ln\langle N\rangle_2^{\rm harm}.
\end{equation}

\noindent
Chemical potentials that go to zero faster than (\ref{val-ca-harm}) for
large $\langle N\rangle_2^{\rm harm}$ vanish in the continuum approximation,
and $(F_2^{\rm harm})_{\mu,\mu}$ diverges. The dominant contributions of the
discrete sum is 

\begin{equation} \label{mu-harm-2D-2}
(F_2^{\rm
    harm})_{\mu,\mu}\simeq\frac{1}{\mu^2}+\frac{1}{\hslash^2\Omega_2^2}\sum_{\substack{n_x,n_y=0\\ (n_x,n_y)\neq(0,0)}}^{{\cal O}\left(\frac{1}{\beta\hslash\Omega_2}\right)}\frac{1}{(n_x+n_y)^2}=\frac{1}{\mu^2}+\frac{\langle N\rangle_2^{\rm harm}}{2\hslash^2\tilde\varrho}\ln\left(\frac{\langle N\rangle_2^{\rm harm}}{\beta^2\hslash^2\tilde\varrho}\right). 
\end{equation}

In one dimension the finiteness of $\tilde\varrho$ prevents the
chemical potential to vanish. Moreover, the constraints
(\ref{val-ca-harm}) and (\ref{num-const-harm}) give the same scaling
for the chemical potential. These scaldings together with the
estimation of (\ref{N-harm1D}) for small chemical potentials imply the
following bound for the density:
$\beta\hslash\tilde\varrho\lesssim\ln\langle N\rangle_1^{\rm
  harm}$. For large $\tilde\varrho$ the explicit equations
(\ref{N-harm1D},\ref{mu-mu-harm1D}) imply  

\begin{equation} \label{mu-harm-1D}
(F_1^{\rm harm})_{\mu,\mu}\simeq\beta\langle N\rangle_1^{\rm harm}\frac{e^{\beta\hslash\tilde\varrho}}{\hslash\tilde\varrho}\lesssim\beta^2\frac{(\langle N\rangle_2^{\rm harm})^2}{\log\langle N\rangle_1^{\rm harm}}..
\end{equation}

\noindent
For extremely high densities, there is a remarkable intermediate
regime between (\ref{val-ca-hom}) and (\ref{num-const-hom}), where the
discrete sums must be evaluated. The dominant contributions of
$\langle N\rangle_1^{\rm harm}$ and $(F_1^{\rm harm})_{\mu,\mu}$ are

\begin{eqnarray} \label{mu-harm-1D-2}
\langle N\rangle_1^{\rm harm} & \simeq & -\frac{1}{\beta\mu}+\frac{1}{\beta\hslash\omega_x}\sum_{n_x=1}^{{\cal O}\left(\frac{1}{\beta\hslash\omega_x}\right)}\frac{1}{n_x}\simeq-\frac{1}{\beta\mu}+\frac{1}{\beta\hslash\omega_x}\ln\left(\frac{1}{\beta\hslash\omega_x}\right), \\
(F_1^{\rm harm})_{\mu,\mu} & \simeq &
\frac{1}{\mu^2}+\frac{1}{\hslash^2\omega_x^2}\sum_{n_x\geqslant 1}\frac{1}{n_x^2}=\frac{1}{\mu^2}+\frac{\pi^2}{6\hslash^2\tilde\varrho^2}\left(\langle N\rangle_1^{\rm harm}\right)^2\lesssim\frac{1}{\mu^2}+\frac{\pi^2\beta^2}{6}\frac{\left(\langle N\rangle_1^{\rm harm}\right)^2}{\ln^2\langle N\rangle_1^{\rm harm}},
\end{eqnarray}

\noindent
where the last inequality holds for densities larger than those in the
continuum approximation, $\beta\hslash\tilde\varrho\gtrsim\ln\langle
N\rangle_1^{\rm harm}$.

\begin{figure}[htbp]
\centering
\includegraphics[width=0.8\columnwidth]{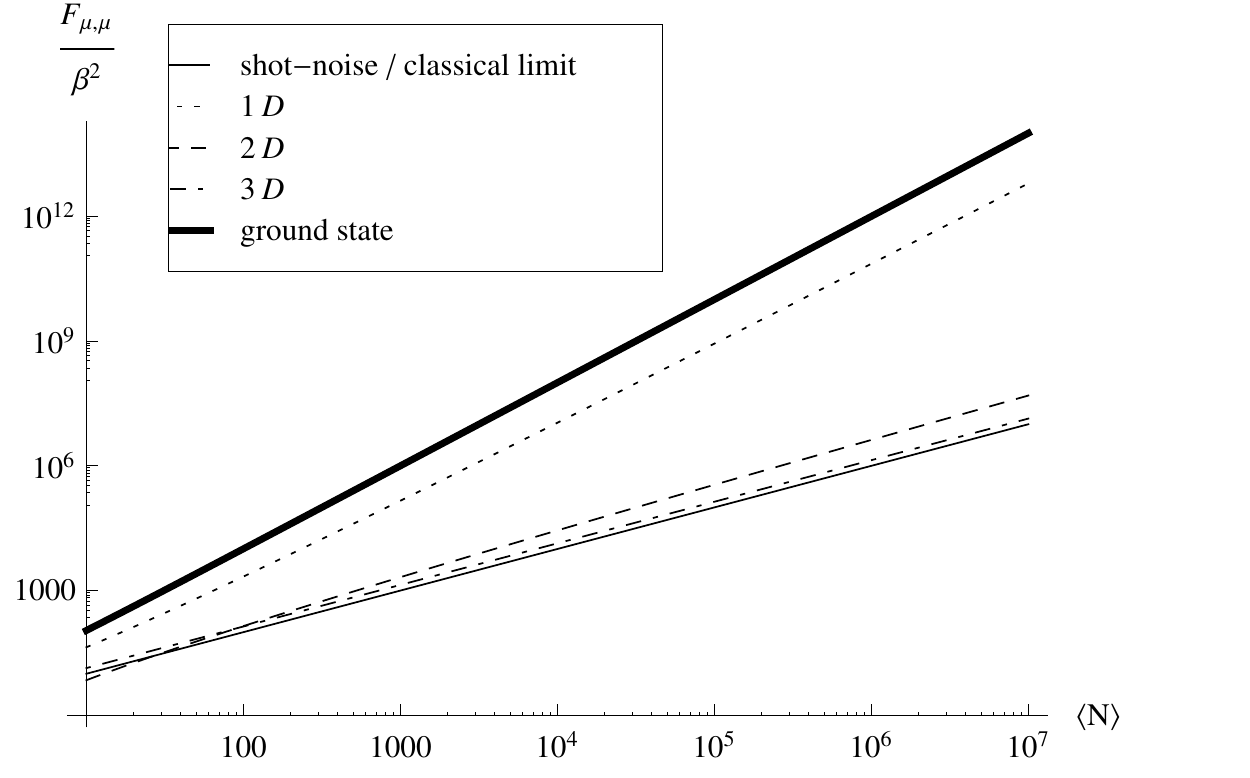}
\caption{Upper bounds of the rescaled Fisher information $(F_d^{\rm
    harm})_{\mu,\mu}/\beta^2$ for harmonically trapped bosons above the
  condensation temperature against the average number of particles in
  double logarithmic scale within the continuum approximation in three
  dimensions (dotdashed line) (\ref{mu-harm-3D}), two dimensions (dashed
  line) (\ref{mu-harm-2D}), and one dimension (dotted line)
  (\ref{mu-harm-1D}). The continuous line is the shot-noise reproduced by
  the classical gas. The thick line is the Fisher information if all particles are in the ground state, i.e.~the zero temperature case.}
\label{harm_gas}
\end{figure}

In Fig. \ref{harm_gas}, we plot the upper bounds
(\ref{mu-harm-3D},\ref{mu-harm-2D},\ref{mu-harm-1D}) of the Fisher
information $(F_d^{\rm harm})_{\mu,\mu}$ versus $\langle N\rangle_d^{\rm
  harm}$ within the continuum approximation in double logarithmic scale. As,
for homogeneous gases, we see that the smaller the dimensionality, the
better the 
sensitivity of the chemical potential. The curves are compared with the
classical limit that exibits shot-noise, i.e. a linear scaling, and the zero
temperature case where all the particles occupy the ground
state. Figs. \ref{hom_gas} and \ref{harm_gas} show that harmonic gases
exhibit worse sensitivities than homogeneous gases with the same dimension,
in accordance with the scalings in the formulas.

\section{Fisher matrix in the presence of Bose-Einstein
  condensation}\label{sec.BEC} 
 
Bosonic gases experience a phase transition towards Bose-Einstein
condensation. This occurs when a macroscopic number of particles
occupy a vanishingly small number of states. These states are either
the only ground state in the case of normal Bose-Einstein
condensation, or a band of states for the so-called generalized
Bose-Einstein condensation. 

The conventional approach to the Bose-Einstein condensation
\cite{Huang,PethickSmith,PitaevkiiStringari,Reichl} consists in
finding the maximum average number of particles within the continuum
limit, i.e. substituting the sum over the modes with an integral. If
the actual number of particles is larger, the continuum limit breaks,
and the sum over the modes has to be replaced by the corresponding
integral plus a singular measure. The latter is a delta-like measure
that singles out the contribution of a zero measure subset of
states. This describes the emergence of the Bose-Einstein condensate
above the critical density $\varrho_c=\max_\mu\varrho$
($\tilde\varrho_c=\max_\mu\tilde\varrho$) or equivalently below the
critical temperature $T_c$ defined by the relation
$\varrho(T_c)=\varrho_c$ ($\tilde\varrho(T_c)=\tilde\varrho_c$). 

Now, we discuss the sensitivity in the estimation of $(\beta,\mu)$ in the
presence of different kinds of Bose-Einstein condensations.

\subsection{Normal Bose-Einstein condensation} \label{normal.BEC}

If the density is finite, homogeneous ideal gases with isotropic confinement
($L_x=L_y=L_z$) exhibit a normal Bose-Einstein condensation only in three
dimensions, with critical temperature $T_c=\frac{2\pi\hslash^2}{k_B
  m}\left(\frac{\varrho}{\zeta(3/2)}\right)^{2/3}$ and fraction of
condensate $f=1-(T/T_c)^{3/2}$
\cite{Huang,PethickSmith,PitaevkiiStringari,Reichl}. Harmonically trapped
ideal gases with the same trap frequency in each direction
($\omega_x=\omega_y=\omega_z$) undergo Bose-Einstein condensation in $d=2,3$
dimensions, with critical temperatures
$T_c=\frac{\hslash}{k_B}\left(\frac{\tilde\varrho}{\zeta(d)}\right)^{1/d}$
and fraction of condensate $f=1-(T/T_c)^d$ \cite{Mullin1997}. We now focus
on the condensed phase of these gases. 

Below the critical temperature $T<T_c$ the chemical potential is very small
at finite size, $-\beta\mu={\cal O}(1/\langle N\rangle)$, and the
contribution of the ground state must be singled out in the sums. The
average number of particles below the critical temperature is $\langle
N\rangle=\langle N_0\rangle+\langle N_{\rm ex}\rangle$, where $N_0=a_0^\dag
a_0$ is the number of particles in the ground state, $\langle
N_0\rangle=(e^{-\beta\mu}-1)^{-1}=f\langle N\rangle$, $N_{\rm ex}=N-N_0$ is the number of
particles in the excited states, and $\langle N_{\rm ex}\rangle$ is given by
(\ref{N-hom}) or (\ref{N-harm}) within the continuum approximation. The
estimation of $\langle N_0\rangle$ at small chemical potentials gives the
evaluation of the chemical potential itself $-\beta\mu\simeq 1/(f\langle
N\rangle)$. Since 
the modes are independent in the grand canonical state, the variance of the
total number of particles is $\Delta^2 N=\Delta^2 N_0+\Delta^2 N_{\rm ex}$,
where $\Delta^2 N_0$ is the variance of the number of particles condensed in
the ground state and $\Delta^2 N_{\rm ex}$ is the variance of the number of
particles in the excited states. The computation at small but finite $\mu$,
that is at finite size, gives $\Delta^2
N_0=e^{\beta\mu}(1-e^{\beta\mu})^{-2}=\langle N_0\rangle+\langle
N_0\rangle^2$. The scaling of the chemical potential violates the conditions
(\ref{val-ca-hom},\ref{val-ca-harm}), thus $\Delta^2 N_{\rm ex}$ must be
evaluated at zero chemical potential and via the computation of the discrete
sum if its continuum approximation diverges at $\mu=0$. The leading
contribution is the same as (\ref{mu-hom-3D-2}), (\ref{mu-harm-3D}) or
(\ref{mu-harm-2D-2}) without the term $1/\mu^2$ and divided by
$\beta^2$. On the other hand, applying the theory of spontaneous symmetry
breaking, one finds that $\Delta^2 N_0=0$, the mode operators of the ground
states being replaced by numbers through the Bogolioubov shift
\cite{Yukalov2005}. 

This apparent ambiguity is sorted out by noticing that the symmetry breaking
approach can be applied only in the thermodynamic limit. In this limit, the
grand canonical thermal state without symmetry breaking is thermodynamically
unstable because the isotermal compressibility (\ref{kappa}) diverges. Thus,
the gas splits up into 
two phases, one consisting of all particles in the ground state without
particle number fluctuations while the other is the non-condensed phase
where statistical averages equal those for the excited states. However, the
grand canonical thermal state can be considered at finite size, even if its
instability grows with the number of particles as argued after equation
(\ref{kappa}). Moreover, the Bogoliubov shift is not always a good
approximation of exact physical behaviours \cite{Gaul2013}. 

The computation of the particle number variance straightforwardly gives the
Fisher information of the chemical potential via (\ref{entries1}). If the
symmetry breaking approach is considered, the variance $\Delta^2 N$ and the
Fisher information reduce to that of the non-condensed phase, namely
(\ref{mu-hom-3D-2}), (\ref{mu-harm-3D}) or (\ref{mu-harm-2D-2}) without the
term $1/\mu^2$. In particular, the Fisher information scales superlinearly
with the number of particles for the three-dimensional homogeneous gas and
the two-dimensional harmonically trapped gas. If the grand canonical thermal
state without the symmetry breaking is considered, the Fisher information
$F_{\mu\mu}$ is increased by the superlinear contribution due to the ground
state,

\begin{equation}
  \label{eq:T0}
\beta^2\Delta^2N_0=\beta^2\left(f\langle N\rangle+f^2\langle
N\rangle^2\right).  
\end{equation}

\noindent
Equation (\ref{eq:T0}) follows directly from the proportionality relation
between the 
Fisher information and the variance of the total number of particles
(\ref{entries1}), and from the above considerations on the variance. 
This quadratic scaling dominates over the
contribution of the excited states, and also the three-dimensional
harmonically trapped gas exhibits sub-shot-noise.   

The contributions of the ground state to the other entries of the Fisher
matrix do not scale superlinearly in the condensed phase: the ground state contribution to $F_{\beta,\beta}$ is $\beta^{-2}$, and the contribution to
$F_{\mu,\beta}$ is $(\beta\mu)^{-1}=-f\langle N\rangle$. Even if the
presence of the BEC does not increase the sensitivity of the temperature
estimation beyond the shot-noise, it is worthwhile to compute such sensitivity in the spirit of a recent experiment \cite{Leanhardt2013}. In
this experiment, a harmonically trapped three-dimensional gas of $^{23}$Na
atoms has been prepared below the condensation critical temperature, and the lowest measured temperature was $450\pm 80$pK. Since even in the condensed
phase the ground state contribution to $F_{\beta,\beta}$ does not dominate,
$F_{\beta,\beta}$ is always represented by the 
thermal averages (\ref{beta-beta-hom},\ref{beta-beta-harm}) taken at
$\beta\mu=0$. These considerations give the
following Fisher information for the homogeneous and harmonic gases in the
condensed phase: 
\begin{equation}
(F_d)_{\beta,\beta}^{\rm
  hom}=\frac{(d^2+2d)\zeta\left(\frac{d}{2}+2\right)m^{\frac{d}{2}}V_d}{2^{2+\frac{d}{2}}\pi^{\frac{d}{2}}\hslash^d\beta^{2+\frac{d}{2}}},
\qquad (F_d)_{\beta,\beta}^{\rm
  harm}=\frac{(d+d^2)\zeta(d+1)}{\hslash^d\beta^{d+2}\Omega_d^d}. \label{eq:IFbb}
\end{equation}

First, we observe that the Fisher information given by (\ref{eq:IFbb}) 
decreases with decreasing temperature the faster
the larger the dimension. Thus, lower dimensions provide
better sensitivities for low temperature thermometry.
Remember that there is a condensed phase at non-zero temperature
only in two and three dimensions for the hamonically trapped gas, and
only in three dimensions for the homogeneous gas. Interestingly,
equations \eqref{eq:IFbb} also provide the leading order of the
expansion around zero temperature for bosonic gases where there is no
phase transition, i.e. no condensed phase at non-zero temperature. In
order to show this, one can invert the equation of state
(\ref{N-hom},\ref{N-harm}) to find the function $\mu(\beta)$, plug it
into (\ref{beta-beta-hom},\ref{beta-beta-harm}), and then perform a
single limit $\beta\to\infty$ using the properties of polylogarithms
\cite{Wood1992}. The inversion of the equation of state can be done
analytically for the two-dimensional homogeneous gas \eqref{N-hom2D}
and the one-dimensional harmonically trapped gas \eqref{N-harm1D}, due
to the simple analytical form, while for the one-dimensional
homogeneous gas it can be done in the limit of small chemical
potential with ${\rm
  Li}_{\alpha}(-e^x)=-x^\alpha/\Gamma(\alpha+1)-\pi^2
x^{\alpha-2}/(6\Gamma[\alpha-1])+{\cal O}(x^{\alpha-4})$ for $Re(x)\gg
1$ \cite{Wood1992}. Intuitively, the fugacity $e^{\beta\mu}$ and the
density (\ref{N-hom},\ref{N-harm}) are infinitesimal in the limit
$\beta\to\infty$, approaching rather the classical limit than the
small temperature limit, unless $\mu\to 0$ such that $\beta\mu\to
0$. This indicates why equations (80) are the leading order of the
small temperature limit in the absence of phase transition, even if
they were derived taking $\mu\to 0$ before $\beta\to\infty$.     

The above $F_{\beta,\beta}$ can be used to derive a lower bound for
the relative error of the temperature estimation. Note that, in the
limit $T\to 0$, the relative error of the optimal estimation
$1/(\beta\sqrt{(F_d)_{\beta,\beta}})\leqslant\sqrt{\rm
  var(\beta)}/\beta=\Delta T/T$ diverges as $1/T^{d/4}$ for
homogeneous gases and $1/T^{d/2}$ for harmonic gases. However, the
optimal relative error is small in actual experiments: for instance,
in the case of the experiment \cite{Leanhardt2013},
i.e. $\omega_x=2\pi (0.65\pm 0.05)$Hz, $\omega_y=2\pi (1.2\pm 0.1)$Hz,
$\omega_z=2\pi (1.81\pm 0.05)$Hz, and $T=450$pK, we get
$1/(\beta\sqrt{(F_3^{\rm harm})_{\beta \beta}})\simeq 0.011$. Our
bound is one order of magnitude smaller than the experimental error
$80/450\simeq 0.18$, suggesting that the realized sensitivity can be
improved.  

\subsection{Normal Bose-Einstein condensation under isobaric cooling}

A different behaviour of bosonic gases occurs when the temperature is
lowered at constant pressure, instead of constant volume, in ideal
homogeneous bosonic gases in two dimensions with vanishing boundary
conditions \cite{Gunther1974}: $\varepsilon_k=(k_x^2+k_y^2)/(2m)$, with 
$k_{x,y}=\frac{\pi}{L}n_{x,y}$ and $n_{x,y}=1,2,3,\dots \,\,$. Under
isobaric cooling, the density diverges, and all the dependences on the
volume must be carefully considered. This explains why this kind of
Bose-Einstein condensation depends on the boundary conditions, because
boundary conditions affect the scaling of the ground state energy with respect to the volume \cite{Gunther1974}. In this 
example, a transition towards a normal Bose-Einstein condensation takes
place at temperature $T_c=\sqrt{\frac{12\hslash P}{\pi m k_B}}$. Equations
(\ref{N-hom},\ref{H-hom},\ref{mu-mu-hom},\ref{beta-beta-hom},\ref{mu-beta-hom}) hold
above the critical temperature. The volume is sub-extensive when it
approaches $T_c$ from above or below $T_c$ (i.e.~the volume scales
sublinearly with the number of particles), and the density diverges. A
divergent particle density is a known mechanism to overcome no-go theorems
\cite{Penrose1956,Hohenberg1967,Krueger1967,Chester1968,Girardeau1969,Chester1969}
for Bose-Einstein condensation in low dimensions
\cite{Sonin1969,Rehr1970,Mullin1997,Mullin2000}. Approaching the critical
temperature, the density is $\varrho=\langle
N\rangle/V_2\simeq-\sqrt{\frac{3mP}{\pi^3\hslash^2}}\ln(T/T_c-1)$, as
computed in \cite{Gunther1974}. Hence $F_{\mu,\mu}$ scales superlinearly
with $\langle N\rangle$ as in equation (\ref{mu-hom-2D}), but not
$F_{\beta,\beta}\simeq \frac{\pi^2}{3\beta^2\lambda_T^2\varrho}\langle
N\rangle$ and $F_{\mu,\beta}\simeq\langle
N\rangle\left(1-\frac{1}{\lambda_T^2\varrho}\right)$. 

Below the critical temperature we need to single out the contribution of the
ground state in the sums. This contribution to $F_{\mu,\mu}$ is
$\beta\Delta^2 N_0=\beta^2(\langle N_0\rangle+\langle N_0\rangle^2)$ which
would imply a quadratic scaling in the Fisher information. However, applying
the theory of spontaneous symmetry breaking, the variance $\Delta^2 N_0$
vanishes because the number of particles in the condensate is approximated
with a number. Moreover, the contribution of the excited states comes from
the continuum approximation of the non-condensed gas. However, the integral
in $F_{\mu,\mu}$ diverges \footnote{Following the analysis of
  \cite{Gunther1974} at finite but large size, one can show that
  substituting the sum with the integral in 
{$F_{\mu,\mu}$ }
gives a
  non-negligible error.}, and we need to compute the discrete sum, whose
leading order comes from the behaviour of small momenta. This contribution is
the same as (\ref{mu-hom-2D-2}) without the term $1/\mu^2$, and scales as
$V_2^2$. Since below the critical temperature the non-extensive scaling of
the volume is $V_2={\cal O}(\sqrt{\langle N\rangle})$, the previous Fisher
information scales linearly in the average number of particles. However, at
the edge of the transition $\langle N\rangle={\cal O}(V_2\ln V_2)$, thus the
Fisher information scales more than linearly with the average number of
particles $F_{\mu,\mu}/\langle N\rangle={\cal O}(V_2/\ln V_2)$. 

As pointed out before, $F_{\beta,\beta}$ and $F_{\mu,\beta}$ scale at most
linearly with the average number of particles both in the  ground state
contribution and in the excited state contribution within the continuum
approximation. 

\subsection{Generalized Bose-Einstein condensation and dimensional confinement}

A different kind of condensation, called generalized Bose-Einstein
condensation, occurs when a band of states of zero measure is
macroscopically 
occupied, rather than only the  ground state. Examples are ideal gases
confined in anisotropic homogeneous or harmonic potentials. If the
confinement is much stronger in some directions, the contributions of the
excited energy levels in the less confined directions dominate below the
critical temperature. In other words, the condensation occurs only in the
ground state of the more confined directions, and an effective lower
dimensional gas is realized. A hierarchy of condensations is possible: from
a three-dimensional gas to a two or one dimensional gas, and from a two
dimensional gas to a one dimensional gas. Generalized Bose-Einstein
condensation has been studied both at finite size and in the thermodynamic
limit focusing on the mathematical structure and general properties of
quantum gases
\cite{Girardeau1960,Girardeau1965,Casimir1968,Krueger1968,Rehr1970,vandenBerg1982,vandenBerg1983,vandenBerg1986,Ketterle1996,vanDruten1997,Mullin1997,Zobay2004,Beau2010,Mullin2012},
in connection with liquid helium in thin films
\cite{Osborne1949,Mills1964,Khorana1964,Goble1966}, magnetic flux of
superconducting rings \cite{Sonin1969}, and gravito-optical traps
\cite{Wallis1996}. Experimental realizations with trapped atoms have been
reported in
\cite{Gorlitz2001,Greiner2001,Esteve2006,vanAmerongen2008,vanAmerongen2008-2,Bouchoule2011,Armijo2011}.
We shall discuss estimation sensitivity of $(\beta,\mu)$ in the presence of
condensation into lower dimensional gases. The scheme is the following:
first prepare a three-dimensional gas in the grand canonical thermal state
with a fixed density, then lower the temperature until the onset of the
generalized condensation. Afterwards, the gas can be employed for the
estimation with sensitivity given by the Fisher matrix. 

For the sake of concreteness, we consider an ideal homogeneous gas confined
in a slab, namely a box of dimension $L_x\times L_y\times L_z$ with
$L_{x,y}\gg L_z$, where condensation in a two-dimensional gas occurs
\cite{Osborne1949,Goble1966,Krueger1968,Sonin1969,vanDruten1997,Beau2010}. This
system is also analytically convenient because the average number of
particles and the Fisher information relative to the chemical potential of
the two-dimensional homogeneous gas have simple expressions
(\ref{N-hom2D},\ref{mu-mu-hom2D}). Note also the formal similarity with the
ideal gas confined in a cigar-like harmonic potential
\cite{Ketterle1996,Mullin1997,vanDruten1997,Beau2010,Mullin2012}, because
both the two-dimensional ideal gas in a box potential and the
one-dimensional ideal gas in a harmonic potential have a constant density of
states. 

The critical density and the critical temperature
of the three-dimensional gas are respectively $\varrho_c^{\rm
  3D}=\zeta(3/2)/\lambda_T^3=\varrho \, (T/T_c^{\rm 3D})^{3/2}$ and
$T_c^{\rm 3D}=\frac{2\pi\hslash^2}{k_B
  m}\left(\frac{\varrho}{\zeta(3/2)}\right)^{2/3}$. If
$\varrho>\varrho_c^{\rm 3D}$ and $T<T_c^{\rm 3D}$, a number of particles
$f\langle N\rangle$, with $f=1-(T/T_c^{\rm 3D})^{3/2}$, condenses in a
small part of the modes. Since $L_{x,y}\gg L_z$, the occupancies $(e^{\beta(\varepsilon_k-\mu)}-1)^{-1}$ of the
energy levels with $n_z\neq 0$ are negligible compared to the others, below
the critical temperature. The remaining modes form a two-dimensional gas
consisting of energy levels with $n_z=0$, and the number of particles
confined there is given by (\ref{N-hom}), namely $\langle
N\rangle_2^{\rm hom}\simeq -\frac{L_x L_y}{\lambda_T^2}\ln(\beta|\mu|)$ for small $\beta|\mu|$. If
these states constitute the condensate then $\langle N\rangle_2^{\rm
  hom}=f\varrho \, L_x L_y L_z$ and $\beta|\mu|\simeq
e^{-f\varrho\lambda_T^2 L_z}$. Moreover, in order for the occupancies of the
modes with $n_z=0$ to contribute with a singular measure in the continuum
limit, the chemical potential should satisfy
$\beta\varepsilon_{(k_x,k_y,0)}\leqslant|\beta\mu|\simeq
e^{-f\varrho\lambda_T^2 L_z}\ll\beta\varepsilon_{(0,0,k_z\neq 0)}$. This
implies that $L_x=L_y\geqslant\gamma e^{\alpha L_z}$ for some constant
$\alpha$ independent of $L_z$ and some function $\gamma(L_z)$ that does not suppress the exponential scaling with $L_z$.

In order to find the behaviour of $\gamma$ 
in the thermodynamic limit,
we now compare the chemical potential with the first excited energy in the
transversal directions, $x$ and $y$,
i.e. $\varepsilon_{(1,0,0)}=\pi\lambda_T^2/(\beta L_x^2)$. The number of
particles in the two-dimensional condensate $\langle 
N\rangle_2^{\rm hom}$  grows when $\beta|\mu|$ decreases, and is estimated in 
the deep two-dimensional condensate phase by its value at
$\beta|\mu|\simeq\beta\varepsilon_{(1,0,0)}$. The latter is the condition
for the possible onset of a second condensation in the ground state alone
where the energy of all the excited states is 
neglible compared to the energy scale
$\beta|\mu|$. If $L_x=L_y\ll\gamma e^{\alpha L_z}$, the gas directly
condenses in the ground state without the intermediate two-dimensional
condensate, and indeed $\beta|\mu|\simeq\beta\varepsilon_{(1,0,0)}$ implies
$\langle 
N\rangle_2^{\rm hom}/V_3\ll
2\alpha/\lambda_T^2-\ln(\pi\lambda_T^2/\gamma^2)/(\lambda_T^2 L_z)$. This
evaluation for the two-dimensional occupation is compatible with the absence
of a two-dimensional condensate, i.e. $\langle 
N\rangle_2^{\rm hom}\ll 1$, provided $(\ln\gamma/\lambda_T)/L_z\to 0$ when $L_z\to\infty$. In the opposite limit $L_x=L_y\gg\gamma e^{\alpha L_z}$, $\langle
N\rangle_2^{\rm hom}/V_3\gg
2\alpha/\lambda_T^2-\ln(\pi\lambda_T^2/\gamma^2)/(\lambda_T^2 L_z)$ implies
that the number of particles in the two-dimensional condensate  grows
indefinitely with decreasing $\beta|\mu|$. In the absence of a saturation, there is no further condensation towards the ground state. Such saturation occurs if $L_x=L_y=\gamma e^{\alpha L_z}$: $\langle
N\rangle_2^{\rm
  hom}/V_3\simeq 2\alpha/\lambda_T^2-\ln(\pi\lambda_T^2/\gamma^2)/(\lambda_T^2
L_z)\to 2\alpha/\lambda_T^2$ in the thermodynamic limit. Thus, there is a
second critical density $\varrho_c^{\rm 2D}=\varrho_c^{\rm
  3D}+2\alpha/\lambda_T^2$: when the density approaches $\varrho_c^{\rm 2D}$
the chemical potential scales as $\beta|\mu|\simeq e^{-2\alpha L_z}$,
whereas if $\varrho>\varrho_c^{\rm 2D}$ a second condensation with a
macroscopic fraction of particles in the ground state occurs and the
chemical potential scales as the inverse of the three-dimensional
volume. 

One can also derive the temperature $T_c^{\rm 2D}$ below which the
occupation of the ground state dominates over the other modes of the
two-dimensional gas, at finite size. This temperature is found by
  imposing that the density equals the second critical density:
  $\rho(T_c^{\rm 2D})=\rho_c^{\rm 2D}$, equivalent to
  $\zeta(3/2)\lambda_{T_c^{\rm 2D}}^{-3}+2\alpha\lambda_{T_c^{\rm
      2D}}^{-2}=\rho$, where $\lambda_{T_c^{\rm 2D}}$ is the thermal
  wavelength evaluated at the second critical temperature \cite{Beau2010}.
As mentioned above, a two-dimensional gas does not condense in the ground
state if the usual thermodynamic limit is considered, with the density fixed
and finite. The reason of such a condensation here is that the number of
particles in the two-dimensional gas is proportional to the total number of
particles of the original three-dimensional gas, thus to the
three-dimensional volume, and the two-dimensional density $\langle
N\rangle_2^{\rm hom}/(L_x L_y)=g\varrho L_z$ diverges in the thermodynamic
limit. We will see that this is also the reason for a superlinear scaling of
the Fisher information as soon as a condensation in a two-dimensional gas
occurs.

We focus on the temperature regime $T_c^{\rm 2D}\leqslant T\leqslant
T_c^{\rm 3D}$, where the number of particles in the two-dimensional gas is
$\langle N\rangle_2^{\rm hom}=\langle N\rangle-\rho_c^{\rm 3D}V_3=f\langle
N\rangle$. The macroscopic occupation of a vanishingly small number of modes, namely the
 two-dimensional gas, below the first critical temperature cannot be
 described by the continuum approximation of the eigenenergies. Thus, the
 contributions of the condensate must be singled out from the integral
 (\ref{cont-hom}) in the thermodynamic averages. For instance, the average
 number of particles is \cite{Beau2010,Mullin2012} 
\begin{equation}
\langle N\rangle=\langle N\rangle_2^{\rm hom}+\langle N\rangle_3^{\rm
  hom}=\frac{V_3}{\lambda_T^2
  L_z}\ln(1-e^{\beta\mu})+\frac{V_3}{\lambda_T^3} \, {\rm
  Li}_{\frac{3}{2}}(e^{\beta\mu})=f\varrho V_3+\frac{1}{\lambda_T^3} \, {\rm Li}_{\frac{3}{2}}(e^{\beta\mu})
\end{equation}

\noindent
Similarly, the entries of the Fisher matrix are the sum of three- and two-dimensional contributions: at the leading orders for large $L_z$,

\begin{eqnarray}
F_{\mu,\mu} & = & (F_2^{\rm hom})_{\mu,\mu}+(F_3^{\rm
  hom})_{\mu,\mu}=\frac{\beta^2 L_x
  L_y}{\lambda_T^2}\left(e^{f\varrho\lambda_T^2
  L_z}-1\right)+\frac{\beta^2 V_3}{\lambda_T^3}{\rm
  Li}_{\frac{1}{2}}(e^{\beta\mu}), \\ 
F_{\beta,\beta} & = & (F_2^{\rm hom})_{\beta,\beta}+(F_3^{\rm hom})_{\beta,\beta} \nonumber \\
& \simeq & \frac{L_x
  L_y}{3\beta^2\lambda_T^2}\left(\pi^2-3e^{-f\varrho\lambda_T^2
  L_z}\right)+\frac{V_3}{\beta^2\lambda_T^3}\left(\beta^2\mu^2{\rm
  Li}_{\frac{1}{2}}(e^{\beta\mu})-3\beta\mu{\rm
  Li}_{\frac{3}{2}}(e^{\beta\mu})+\frac{15}{4}{\rm
  Li}_{\frac{5}{2}}(e^{\beta\mu})\right), \\ 
F_{\mu,\beta} & = & (F_2^{\rm hom})_{\mu,\beta}+(F_3^{\rm
  hom})_{\mu,\beta} \simeq \frac{L_x
  L_y}{\lambda_T^2}\left(-1+\frac{e^{-f\varrho\lambda_T^2
    L_z}}{2}\right)-f\varrho \,
V_3+\frac{V_3}{\lambda_T^3}\left(\beta\mu{\rm
  Li}_{\frac{1}{2}}(e^{\beta\mu})-\frac{3}{2}{\rm
  Li}_{\frac{3}{2}}(e^{\beta\mu})\right). 
\end{eqnarray}

\noindent
The entries $F_{\beta,\beta}$ and $F_{\mu,\beta}$ are linear in the volume, thus in the average number of particles $\langle N\rangle$. On the other hand

\begin{equation}
\frac{F_{\mu,\mu}}{\langle N\rangle}=\frac{\beta^2}{\varrho\lambda_T^2
  L_z}\left(e^{f\varrho\lambda_T^2
  L_z}-1\right)+\frac{\beta^2}{\varrho\lambda_T^3}{\rm
  Li}_{\frac{1}{2}}(e^{\beta\mu}). 
\end{equation}

The two contributions, coming respectively from the two-dimensional
condensate and the three-dimensional bulk, diverge as 
$L_z\to\infty$. To see the divergence of the contribution from the
three-dimensional cloud, we recall that the chemical potential between the
two critical temperatures satisfies $\beta|\mu|\simeq
e^{-f\varrho\lambda_T^2 L_z}$. This value is larger than the minimum energy
spacing, that is $\beta(2\pi\hslash)^2/(2m\gamma^2 e^{2\alpha L_z})$, thus
the continuum approximation still holds in analogy to the discussion of the
bound (\ref{cont-hom}). Therefore, for small chemical potentials the
approximation ${\rm Li}_{1/2}(e^{\beta\mu})=\sqrt{\pi/(\beta|\mu|)}+{\cal
  O}(1)$ implies that the last term of $F_{\mu,\mu}/\langle N\rangle$
behaves as 
$\frac{\sqrt{\pi}\beta^2}{\varrho\lambda_T^3} \, e^{f\varrho\lambda_T^2 L_z/2}$.

In particular, if $L_x=L_y=\gamma e^{\alpha L_z}$, then $\gamma^2 L_z e^{2\alpha L_z}=V_3=\langle N\rangle/\varrho$ and $2\alpha  L_z\simeq\ln (V_3/\ell^3)$, for large $L_z/\ell$ where $\ell$ is a characteristic length, e.g. $\ell=\lambda_T$ or $\ell=1/\alpha$. Thus, 

\begin{equation} \label{fisher_gen_bec}
\frac{F_{\mu,\mu}}{\langle
  N\rangle}=\frac{2\alpha\beta^2}{\varrho\lambda_T^2\ln\frac{\langle
    N\rangle}{\varrho \, \ell^3}}\left(\left(\frac{\langle
  N\rangle}{\varrho \, \ell^3}\right)^{\frac{f\varrho\lambda_T^2}{2\alpha}}-1\right)+\frac{\sqrt{\pi}\beta^2}{\varrho\lambda_T^3}\left(\frac{\langle N\rangle}{\varrho \, \ell^3}\right)^{\frac{f\varrho\lambda_T^2}{4\alpha}}.
\end{equation}

\begin{figure}[htbp]
\centering
\includegraphics[width=0.8\columnwidth]{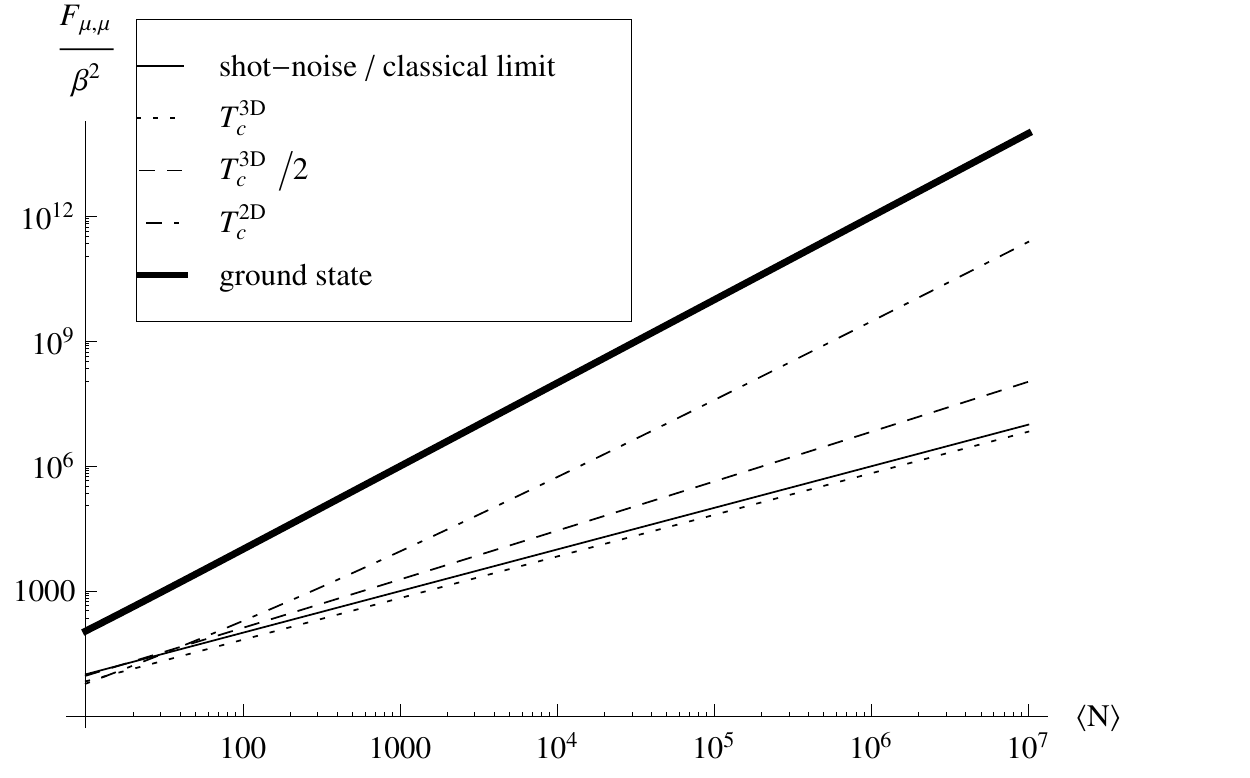}
\caption{Double logarithmic plot of the rescaled Fisher information
  $F_{\mu,\mu}/\beta^2$ of an ideal bose gas versus the average number of particles of a gas of
  $^{87}$Rb atoms in a slab with exponential anisotropy $L_{x,y}\sim\gamma
  e^{\alpha L_z}$. The curves refer to different temperatures: classical
  limit (continuous line, linear in $\langle N\rangle$), first critical
  temperature $T_c^{\rm 3D}=100$nK 
  (dotted line), $T_c^{\rm 3D}/2$ (dashed line), second critical temperature
  $T_c^{\rm 2D}=20$nK (dotdashed line), zero temperature (thick continuous
  line, quadratic in $\langle N \rangle$ (see eq.(\ref{eq:T0}))). The two
  critical temperatures are uniquely determined by $\rho=13\cdot
  10^{12}$cm$^{-3}$ and $\alpha=10\mu$m$^{-1}$. Furthermore,
  $\ell=\lambda_T$.} 
\label{gen_bec}
\end{figure}

Fig. \ref{gen_bec} is the log-log plot of the Fisher information in equation
(\ref{fisher_gen_bec}) against the average number of particles at different
temperatures between the two critical temperatures. The slopes show
inceasing superlinear scalings comprised between the classical limit
exhibiting a linear scaling and the zero temperature case, i.e. all
particles in the ground state. Below the second critical temperature, a
macroscopic number of particles $\langle N_0\rangle$ occupy the ground
state. The contribution of this second BEC must be singled  out from
statistical averages. As discussed in section \ref{normal.BEC}, its
contribution to the Fisher information $F_{\mu,\mu}$ is $\beta^2(\langle
N_0\rangle+\langle N_0\rangle^2)$. Thus, a quadratic scaling emerges with an
increasing weight when temperature decreases.

We emphasize that the one-dimensional ideal gas trapped in a harmonic
potential has mathematically similar properties as the two dimensional ideal
gas in an infinite square potential. In particular, a three-dimensional
harmonically trapped gas with frequencies $\omega_{x,y,z}$ in the three
directions can be 
confined to the $x$ direction if $\omega_{x}\ll\omega_{y,z}$, as studied in 
\cite{Ketterle1996,Mullin1997,vanDruten1997,Beau2010,Mullin2012}. The
equations for the average number of particles and the Fisher matrix in the
condensed phase are the sum of three- and one-dimensional contributions,
similarly to the above discussed case of homogeneous gas. For instance,

 \begin{equation} \label{fisher1Dharm}
\frac{F_{\mu,\mu}}{\langle N\rangle}=\frac{\beta\omega_y\omega_z}{\tilde\varrho\hslash}\left(e^{\frac{\hslash\beta\tilde\varrho}{\omega_y\omega_z}}-1\right)+\frac{{\rm Li}_2(e^{\beta\mu})}{\tilde\varrho\beta\hslash^3}, 
\end{equation}

\noindent
which diverges in the thermodynamic limit. Notice that now the contribution
from the three-dimensional harmonically trapped gas ${\rm Li}_2(e^{\beta\mu})$ is always bounded, unlike the analogous contribution in the case of the square
potential.

\section{Fisher matrix with interacting Bose gases} \label{sec.int}

In this section we discuss how the interactions modify the aforementioned
superlinear scaling of the Fisher information $F_{\mu,\mu}$. It is already
known that small interactions can wipe out the superlinear scaling of
$F_{\mu,\mu}$, for the three-dimensional homogeneous gas below the critical
temperature \cite{Huang,Yukalov2005}. We now consider three models of
interactions which have different effects on the Fisher information. The
first model describes harmonic interactions which preserve the superlinear
scalings discussed above. Then, we discuss mean field interactions for
  which a complete analytic solution can be computed, resulting in the
  suppression of the superlinear scaling of the Fisher information unless
  the interaction strength is very small. Finally, we consider contact
  interactions which do not allow a general analytic computation but can be
  treated perturbatively for small interactions. This regime, that is
  experimentally accessible \cite{Armijo2011}, is characterized by only
  small deviations from sub-shot-noise. In the latter two models, there is
a tradeoff between the smallness of the interaction and the strength of the
sub-shot-noise: the stronger the gain over the shot-noise, the weaker the
interaction must be in order to preserve sub-shot-noise. 

\subsection{Harmonic interactions}
A simple model for interacting systems is given by harmonic
interactions \cite{Madelung08}. Consider for the moment the interacting hamiltonian at
the level of first quantization, e.g. 

\begin{equation}
H_{\rm 1st}=\sum_j \left(\frac{{\bf p}_j^2}{2m}+\frac{m\omega^2 {\bf r}_j^2}{2}\right)+\sum_{j,l}\gamma ({\bf r}_j-{\bf r}_l)^2,
\end{equation}

\noindent
where ${\bf p}_j$ and ${\bf r}_j$ are respectively the momentum and the position of the $j$-th particle. The hamiltonian is a quadratic form in the variables
$\{{\bf p}_j,{\bf r}_j\}_j$. Therefore, via a rotation in the phase-space $(\tilde{\bf p}_1,\tilde{\bf r}_1,\tilde{\bf p}_2,\tilde{\bf r}_2,\dots)=R({\bf p}_1,{\bf r}_1,{\bf p}_2,{\bf r}_2,\dots)$ where
$R$ is an orthogonal matrix, the hamiltonian can be recast into a
non-interacting-like hamiltonian $\tilde H_{\rm 1st}=\sum_j
\left(\frac{\tilde{\bf p}_j^2}{2\tilde m}+\frac{\tilde m\tilde\omega^2\tilde{\bf r}_j^2}{2}\right)$. Moving to the second quantization, we have a
problem formally similar  to that of an ideal gas in a harmonic
potential, discussed in the previous sections. The interactions in
this model are hidden in the phase-space rotation which maps single
particle operators and modes into collective operators and
modes. Thus, the above considerations on harmonically trapped gases
apply with the substitution $\omega_{x,y,z}\to\tilde\omega_{x,y,z}$. 

\subsection{Mean field interaction}
Here, we focus on the Bose gas with mean field interaction, also known
as imperfect Bose gas. The hamiltonian is $H=H_0+\lambda N^2/(2V_d)$,
where $H_0$ is the hamiltonian of the ideal gas as in (\ref{ham}), $N$
is the total number operator (\ref{numb}), $V_d$ is the volume, and
$\lambda$ is the interaction strength. $\lambda$ is positive for
repulsive interactions, that always take place at small
distances. This statistical model has been solved in
\cite{vandenBerg1984}, for a general class of non-interacting
hamiltonians independently of the dimensionality. The grand canonical
thermodynamic potential, i.e. the pressure, is: 

\begin{equation} \label{gc.pot}
p_\lambda=\frac{1}{V_d\beta}\ln Z_G^{(\lambda)}=\frac{(\mu-\alpha(\mu))^2}{2\lambda}+p_0(\alpha(\mu)),
\end{equation}

\noindent
where $Z_G^{(\lambda)}$ is the grand canonical partition function of
the mean field model, $p_0=\lim_{\lambda\to 0}p_\lambda$ is the
pressure of the non interacting gas, $\alpha(\mu)$ is zero if
$\mu\geqslant\lambda\rho_c$ and is the unique solution of
$\alpha+\lambda \, \partial_\alpha p_0(\alpha)=\mu$ if
$\mu<\lambda\rho_c$, and $\rho_c$ if the critical density which
coincides with that of the non-interacting hamiltonian. From the
thermodynamic potential we can compute all the statistical averages,
for instance 

\begin{eqnarray}
\langle N\rangle_\lambda & = & \frac{1}{\beta}\frac{\partial}{\partial\mu}\ln Z_G^{(\lambda)}=\frac{V_d}{\lambda}(\mu-\alpha(\mu)), \\
\Delta^2_\lambda N & = & \frac{1}{\beta^2}\frac{\partial^2}{\partial\mu^2}\ln Z_G^{(\lambda)}=\frac{V_d}{\beta}\frac{\partial^2_\alpha p_0(\alpha)}{1+\lambda \, \partial^2_\alpha p_0(\alpha)}=\frac{V_d\Delta^2_0 N}{V_d+\lambda\beta\Delta^2_0 N}, \\
F_{\mu,\mu}^{(\lambda)} & = & \beta^2\Delta^2_\lambda N=\frac{V_d\beta F_{\mu,\mu}^{(0)}}{V_d\beta+\lambda F_{\mu,\mu}^{(0)}}.
\end{eqnarray}

\noindent
From these computations, we learn that the imperfect Bose gas has the same
critical density as the ideal gas with the same non-interacting
hamiltonian. In particular, there is the same hierarchy of condensation
mentioned in the previous section: a three-dimensional gas condenses into a
two-dimensional gas which may condense in the unique ground state. However,
the thermodynamic properties differ from those of the ideal gas. For
instance, the Fisher information $F_{\mu,\mu}^{(\lambda)}$ always scales at
most linearly with the volume, for finite interaction strengths. Indeed,
even if $F_{\mu,\mu}^{(0)}$ scales superlinearly with the average number of
particles, then $F_{\mu,\mu}^{(\lambda)}\simeq V_d\beta/\lambda$, i.e.~is simply extensive due to the proportionality of the volume $V_d$ to the
total particle number, unless
$\lambda\ll V_d\beta/F_{\mu,\mu}^{(0)}$. This happens both in the
three-dimensional bulk and in the lower dimensional condensate.

If the ideal system exhibits a superlinear scaling of the Fisher
information, the bound $\lambda\ll V_d\beta/F_{\mu,\mu}^{(0)}$ goes to
zero for infinite size. In this limit, the superlinear scaling
disappears for any coupling constant, but at finite size there are
values of $\lambda$ which do not destroy the sub-shot-noise. 

\subsection{Contact interaction}

From a theoretical perspective, the study of Bose-Einstein
condensation in statistical mechanics becomes a highly non-trivial
problem in the presence of interactions, see \cite{Zagrebnov2001} for
a review. In particular, the condensation is driven not only by the
decreasing of temperature but also by the presence of
interactions. This implies the coexistence of different condensates
and a complex structure of the states occupied in the condensed
phases, without a clear extension of the generalized Bose-Einstein
condensation in lower dimensional gases. On the other hand, a
kinematic approach was investigated to prove dimensional confinement
\cite{Olshanii1998}, later implemented in mesoscopic systems
\cite{Gorlitz2001,Greiner2001,Esteve2006,vanAmerongen2008,vanAmerongen2008-2,Bouchoule2011,Armijo2011}.
This approach consists in proving that the scattering amplitudes of a
three-dimensional bosonic gas with contact interactions and a strong
harmonic confinement in two transverse dimensions correspond to those
of an effective one-dimensional gas, if the incident wave is frozen in
the transverse ground state and its longitudinal kinetic energy is
smaller than the energy spacing of the transverse potential. The
effective one dimensional gas is described by the Yang-Yang model
which was formally solve in \cite{Yang1968}. 

The hamiltonian of the model is

\begin{equation}
H=\sum_k\frac{k^2}{2m}a_k^\dag a_k+\frac{c}{2L_x}\sum_{k_1,k_2,q}a_{k_1-q}^\dag a_{k_2+q}^\dag a_{k_2}a_{k_1},
\end{equation}

\noindent
and its statistical properties are relevant in a number of experiments
that realize this system
\cite{Gorlitz2001,Esteve2006,vanAmerongen2008,vanAmerongen2008-2,Bouchoule2011,Armijo2011}.
The peculiarity of this statistical model is that the excitations at
infinite interaction strength behave as non-interacting fermions,
while of course they are single-particle bosonic modes for vanishing
interactions. It is desirable for applications to know the Fisher
matrix in the presence of contact interaction. 

The formal computation of statistical averages involves the solution
of two coupled nonlinear integral equations, which can be solved only
numerically. Nevertheless, some analytical results were derived, such
as the second order coherence function $g^{(2)}$ in different regimes
perturbatively for weak and strong interactions \cite{Deuar2009}. Thus,
we focus on the Fisher information of the chemical potential
$F_{\mu,\mu}$ which shows sub-shot-noise in the limit of zero
interaction and for high densities or fixed volumes
(\ref{mu-hom-1D}). Given the second order coherence function, we can
compute the variance of the total number of particles, and then the
Fisher information $F_{\mu,\mu}$ (\ref{entries1.gen}). The relation
between the second order coherence function and the total number
variance is reported in (\ref{var-g2}) and proved in appendix
\ref{g2}. 

Different regimes are parametrized by the two dimensionless quantities,
$\gamma=\frac{2\pi\beta c}{\lambda_T^2\varrho}$ and
$\tau=\frac{4\pi}{\lambda_T^2\varrho^2}$. The $g^{(2)}$ function for strong
interactions, $\gamma\gg 1$, exhibits the typical fermionic anti-bunching
behaviour, namely $0\leqslant g^{(2)}\leqslant 1$ \cite{Deuar2009}. This
property causes a reduction in the variance $\Delta^2N$, and thus of
$F_{\mu,\mu}$, with respect to the shot-noise, as is clear from equation
(\ref{var-g2}). On the other hand, bunching $g^{(2)}>1$, typical of
non-interacting bosons, is responsible of a superlinear scaling of
$\Delta^2N$ and $F_{\mu,\mu}$ with $\langle N\rangle$. The quantum
degenerate gas with small interactions, $\sqrt{\gamma}\ll\tau\ll 1$, is
close to the ideal bosonic gas. Therefore, the $g^{(2)}$ function was
derived in \cite{Deuar2009} within perturbation theory in the
coupling constant $c$: 

\begin{equation}
g^{(2)}(r)=1+\left(1-\frac{4\gamma}{\tau^2}\left(1+\varrho\tau r\right)\right)e^{-\varrho\tau r}.
\end{equation}

Plugging this formula into equation (\ref{var-g2}), we get the Fisher information

\begin{eqnarray}
\label{mu-mu-int}
F_{\mu,\mu} & = & \beta^2\langle
N\rangle+\frac{\beta^2\lambda_T^2\varrho^2}{2\pi}\langle
N\rangle-\frac{\beta^2\lambda_T^4\varrho^4}{8\pi^2}\left(1-e^{-\frac{4\pi}{\lambda_T^2\varrho^2}\langle 
  N\rangle}\right) \nonumber \\ 
& &
+c\left(\frac{3\beta^3\lambda_T^6\varrho^7}{16\pi^3}\left(1-e^{-\frac{4\pi}{\lambda_T^2\varrho^2}\langle 
  N\rangle}\right)-\frac{\beta^3\lambda_T^4\varrho^5}{4\pi^2}\langle
N\rangle\left(2+e^{-\frac{4\pi}{\lambda_T^2\varrho^2}\langle
  N\rangle}\right)\right). \label{fisher_int}
\end{eqnarray}

\noindent
The condition $\sqrt{\gamma}\ll\tau$ is equivalent to
$c\ll\frac{8\pi}{\beta\lambda_T^2\varrho^3}$, and $\tau\ll 1$ reads
$\lambda_T^2\varrho^2\gg 4\pi$. Notice that the Fisher information
$F_{\mu,\mu}$ scales linearly with $\langle N\rangle$, if the density
$\varrho=\langle N\rangle/L_x$ is fixed. Instead, if the size $L_x$ is
fixed, superlinear scaling emerges. 

Let us consider the first line of (\ref{mu-mu-int}), that is the Fisher
information of the ideal gas. The second contribution in the first line of
(\ref{mu-mu-int}), $\frac{\beta^2\lambda_T^2}{2\pi L_x^2}\langle
N\rangle^3$, is exactly the Fisher information already found for the
one-dimensional ideal gas with small chemical potentials, hence large number
of particles. It could seem that the third contribution in the first line of
(\ref{mu-mu-int}) scales quartically with $\langle N\rangle$, for fixed
$L_x$, and thus dominates. This is impossible, since this contribution is
negative and the Fisher information is non-negative by definition. However,
condition (\ref{num-const-hom}) and equation (\ref{N-hom}) for small
chemical potentials imply $\lambda_T^2\varrho^2\lesssim\pi\langle
N\rangle$. Thus, for very large numbers of particles $\langle
N\rangle\gg\frac{\lambda_T^2\varrho^2}{4\pi}\gg 1$, and the absolute value
of the third term is therefore much smaller than the second
one. 

Now we consider the second line of (\ref{mu-mu-int}), namely the
corrections to the ideal gas due to small interactions. These
contributions are linear in $\langle N\rangle$ if the density is
fixed, but they scale superlinearly if the size $L_x$ is fixed. The
last term in (\ref{mu-mu-int}) dominates for $\langle
N\rangle\gg\frac{\lambda_T^2\varrho^2}{4\pi}\gg 1$, and the leading
order of the correction due to the interactions is negative, and its
absolute value is much smaller than the leading order without
interactions: 

\begin{equation}
-c \, \frac{\beta^3\lambda_T^4}{2\pi^2 L_x^5}\langle N\rangle^6\ll -\frac{4\beta^2\lambda_T^2}{\pi L_x^2}\langle N\rangle^3.
\end{equation}

\noindent
The inequality is a consequence of the condition
$\sqrt{\gamma}\ll\tau\Leftrightarrow
c\ll\frac{8\pi}{\beta\lambda_T^2\varrho^3}$. Since the corrections to
the Fisher information are negative, interactions counteract the
sub-shot-noise. However, the corrections are much smaller that the
Fisher information without interactions. Hence, the sub-shot-noise of
one-dimensional homogeneous ideal gases is robust against the
detrimental effect of small contact interactions
$c\ll\frac{8\pi}{\beta\lambda_T^2\varrho^3}$. Sub-shot-noise can be
observed at fixed volume or high density. This condition reduces the
range of admissible coupling constants. Moreover, the condition
$\tau\ll 1$ implies $\lambda_T^2\varrho^2\ll 4\pi$, thus $c\ll
2/\beta$. It is also interesting that the bound
$\frac{8\pi}{\beta\lambda_T^2\varrho^3}$ increases with the
temperature. Although (\ref{mu-mu-int}) is the first perturbative
order, a superlinear scaling of the particle fluctuations, thus of the
Fisher information, was experimentally observed at fixed volume even
beyond the condition $\sqrt{\gamma}\ll\tau$ \cite{Armijo2011}.

\begin{figure}[htbp]
\centering
\includegraphics[width=0.8\columnwidth]{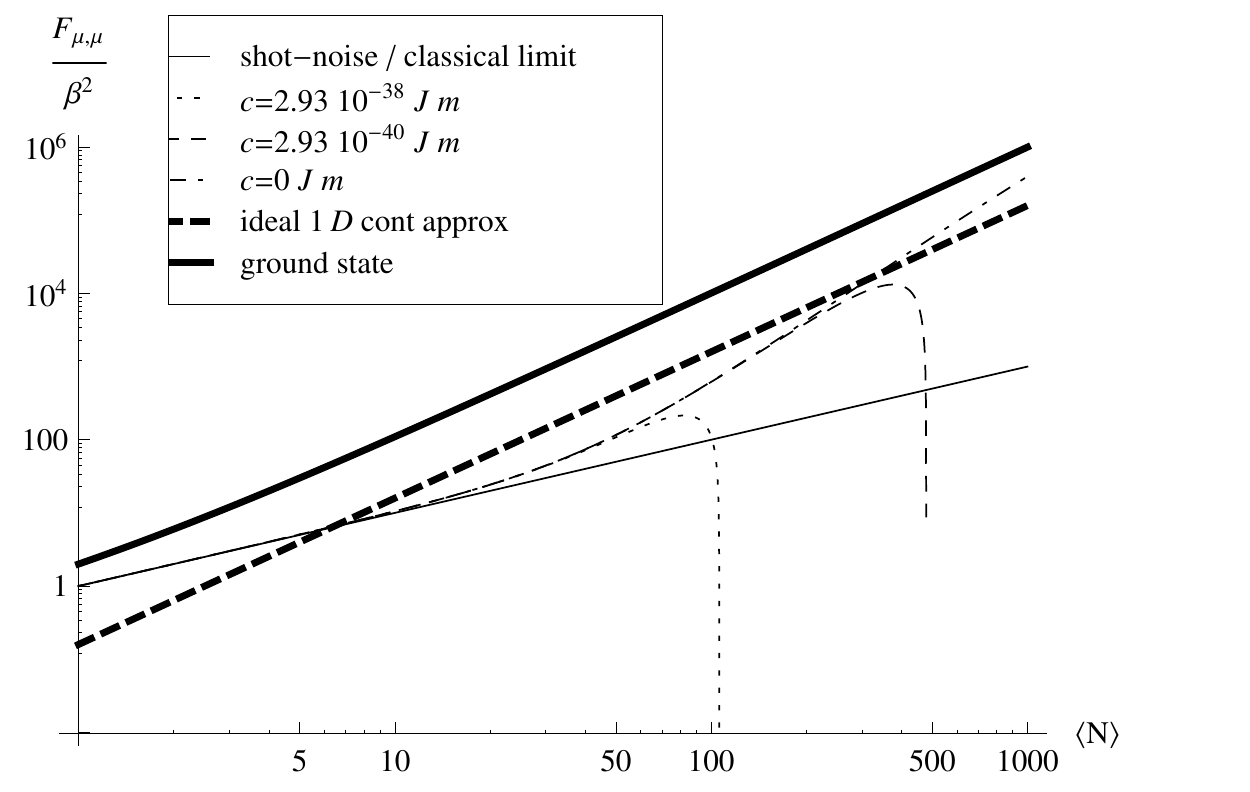}
\caption{Log-log plot of the rescaled Fisher information
  $F_{\mu,\mu}/\beta^2$ in the perturbative regime (\ref{fisher_int})
  against the average number of particles. The results are for a gas of
  $^{87}$Rb 
  atoms confined in $L_x=4.5\,\mu$m at $T=510\,$nK and interaction strength
  $c=2.93 \cdot 10^{-40}$ J m (dotted line), $c=2.93 \cdot 10^{-38}$ J m
  (dashed line), and $c=0$ (dotdashed line). The continuous line is the
  shot-noise resulting from the classical limit, the thick dashed line is
  the best scaling achievable within the continuum approximation
  (\ref{mu-hom-1D}), and the thick line is the Fisher information if all
  particles are in the ground state, i.e.~the zero temperature case.} 
\label{cont_int}
\end{figure}

In Fig. \ref{cont_int}, we plot the perturbative formula (\ref{fisher_int})
at different interaction strengths in double logarithmic scale. The sudden
drop in the curves is a signature of the failure of the perturbative
expansion, where additional terms are required. First, we notice that
decreasing the interaction, the superlinear regime is observed for larger
average number of particles. Moreover, we considered $^{87}$Rb atoms
confined in a fixed size $L_x=4.5\,\mu$m at temperature $T=510\,$nK. These are
the parameters of a recent  experiment where superlinear particle
number fluctuations were observed even beyond the perturbative regime close
to the non-interacting case \cite{Armijo2011}. Our results are in agreement
with the experimental data, considering that the Fisher information of the
chemical potential is proportional to the particle number fluctuations, and
taking into account the experimental sensitivity of the camera and the
depletion of  particles in the transversal excited states as explained in the article
\cite{Armijo2011}.
Furthermore, this experiment
implies that it is possible to check the quantum sensitivity for the
estimation of the chemical potential in one-dimensional interacting gases.

\section{Conclusions} \label{sec.concl}
In summary, we have given a detailed investigation of the sensitivity
with which temperature and chemical potential of quantum gases can be
measured. This was done by calculating the Quantum Fisher information
matrix first for ideal fermionic and bosonic gases, and then examining
three different models of interacting gases. In agreement with
previously known results we have shown that the best sensitivity of
temperature measurements of ideal quantum gases shows SQL-like scaling with the number of particles, both for fermionic and bosonic gases, and irrespective of whether or not the latter are close to the condensation transition. As function of temperature, the relative error diverges as $1/\sqrt{T}$ for homogeneous and harmonically trapped fermionic gases, as $1/T^{d/4}$ for homogenesous BECs, and as $1/T^{d/2}$ for harmonically trapped BECs. This demonstrates that in addition of the impossibility of reaching absolute zero temperature according to the third law of thermodynamics, it also becomes increasingly difficult to measure how close to absolute zero temperature one is. The relative uncertainty increases more rapidly for bosons than for fermions at small temperatures and dimensions larger than one, reflecting the bunching behavior of the bosons.

The sensitivity for
measurements of the chemical potential, which has immediate
applications to the ultimate precision of voltage measurements in
electrical conductors, has a richer behaviour.  While for fermions the
SQL, corresponding to a linear scaling of the quantum Fisher
information with the particle number $N$, cannot be surpassed, bosonic
gases allow in principle enhanced sensitivity beyond the SQL. We have
shown this in different scenarios: standard isochoric BEC; isobaric
BEC; and generalized BEC in 2D or 1D samples, where a hierarchy of
condensation transitions can arise, with condensation first taking
place in a subspace of Hilbert space. The superlinear scaling of the
Fisher information relative to the chemical potential originates in
the macroscopic occupation and the consequent bunching induced
fluctuations in the occupation of the eigenstates of the
BEC. Furthermore, the superlinear scaling never beats the
fluctuations at $T=0$ that scale quadratically with $N$. These
results are not modified in a simple model of harmonic
interactions. However, in a model of mean-field interactions the
super-linear scaling of the quantum Fisher information is destroyed,
unless the interactions are very small. Small contact interactions,
treated in perturbation theory, lead to small corrections of the
super-linear scaling of the quantum Fisher information in
one-dimensional quantum degenerate gas, and indicate that the chemical
potential of bosons can indeed be measured with sub-SQL sensitivity. 

We stress that the two parameters can be jointly estimated. This
is an unusual feature of quantum multivariate estimation, and stems
from the fact that our problem can be recast as a classical
statistical problem in the representation of eigenstates of the energy
and the particle number. The optimal joint estimation consists in
measuring $\mu N-H$ and $N$, while the optimal single parameter
estimation of temperature (chemical potential) can be achieved via the
measurement of $\mu N-H$ ($N$). The number of particles is measured
with absorption imaging in a number of experiments
\cite{Muller2010,Sanner2010,Armijo2011}, while the temperature
estimation and the corresponding optimal measurement of $\mu N-H$ is
hard to implement. The experimental observation of particle number
fluctuations in the light of the present analysis opens the
possibility for future realizations of quantum sensors based on the
measurement of the chemical potential. Furthermore, superlinear
sensitivities can be achieved by different physical systems, such as
dipolar BEC where superpoissonian particle number fluctuations have
been numerically computed \cite{Bisset2013}. \\ 

{\em Acknowledgement} U.M. acknowledges funding by Centre national de la recherche scientifique, by Deutsche Forschungsgemeinschaft, and by Evaluierter Fonds der Albert Ludwigs-Universitaet Freiburg.

\appendix

\section{Classical gases and quantum gases of distinguishable particles}
\label{class.lim}

In this appendix, we derive the Fisher matrix for classical ideal
gases and quantum ideal gases of distinguishable particles. The latter
case is equivalent to the former: once the grand canonical thermal
state of the quantum gas is written in its eigenbasis in first
quantization, the computation of the partition function and all the
statistical averages is same as the for the classical gas. Indeed,
distinguishable particles fulfill the Maxwell-Boltzmann counting and
they both suffer from the Gibbs' paradox. Thus, we shall discuss only
classical gases. 

The hamiltonian of $N$ particles of an ideal gas is the sum of single
particles hamiltonians $H=\sum_{j=1}^N H_j$, where $H_j$ is the
hamiltonian of the $j$-th particle. Consider a gas of identical
classical particles, $H_j=H_1$ for all $j$. The grand canonical
partition function is 

\begin{equation}
Z_G=\sum_{N=0}^\infty \frac{(e^{\beta\mu}Z_c)^N}{N!}=e^{e^{\beta\mu}Z_C},
\end{equation}

\noindent
where $Z_C=\sum_{H_1}e^{-\beta H_1}$ is the canonical partition
function of the single particle problem, and the factor $N!$ is the
cure to the Gibbs' paradox \cite{Huang,Reichl}. This factor arises
naturally in the classical limit of quantum gases, i.e. for high
temperature and low density equivalent to $e^{\beta\mu}\ll 1$
\cite{Huang}. Define the single particle mean energy $\langle
H_1\rangle=-\partial\ln Z_C/\partial\beta$ and its variance $\Delta^2
H_1=\partial^2\ln Z_C/\partial\beta^2$. From the partition function,
we compute the average number of particles, the average total energy,
and the Fisher matrix: 

\begin{eqnarray}
\label{class-av} & & \langle N\rangle=e^{\beta\mu}Z_C, \qquad \langle
H\rangle=\langle N\rangle(\langle H_1\rangle-\mu), \\ 
\label{class-mu-mu} & & F_{\mu,\mu}=\beta^2\Delta^2 N=\beta^2\langle
N\rangle, \\ 
\label{class-beta-beta} & & F_{\beta,\beta}=\Delta^2 (\mu N-H)=\langle
N\rangle((\mu-\langle H_1\rangle)^2+\Delta^2 H_1), \\ 
\label{class-mu-beta} & & F_{\mu,\beta}=F_{\beta,\mu}=\beta \, {\rm
  Cov}(N,\mu N-H)=\beta\langle N\rangle(\mu-\langle H_1\rangle). 
\end{eqnarray}

\noindent
Notice that the entries of the Fisher matrix scale lineraly with the
average number of particles, in accordance with the shot-noise. 

For the homogeneous gas in $d$ dimensions, the single particle
hamiltonian is $H_1=\frac{p^2}{2m}$ which gives $Z_C=V_d(2\pi
m/\beta)^{d/2}$, $\langle H_1\rangle=d/(2\beta)$, and $\Delta^2
H_1=d/(2\beta^2)$. For the $d$-dimensional gas in a harmonic
potential, the single particle hamiltonian is
$H_1=\frac{p^2}{2m}+\frac{1}{2}m\omega^2x^2$ which gives $Z_C=(2\pi
m/\beta)^{d/2}$, $\langle H_1\rangle=d/\beta$, and $\Delta^2
H_1=d/\beta^2$. For quantum gases of distinguishable particles all
with the same single particle hamiltonian, one has to consider the sum
over the eigenvalues of $H_1$ in $Z_C$. The computations provide
prefactors depending on $\hslash$, which do 
 not affect statistical
averages with respect to classical gases. 

\section{Estimation problem for ideal fermionic gases at zero temperatures}
\label{0temp}

One of the assumptions for the validity of the quantum Cram\'er-Rao
bound is the differentiability of the density matrix, as is evident
from the formula of the Fisher matrix (\ref{fisher.gen}). If this
assumption is not met, the quantum Cram\'er-Rao bound is generalized by
the quantum Hammerseley-Chapman-Robbins-Kshirsagar bound
\cite{Tsuda2005}. The Fermi-Dirac distribution involved in the grand
canonical thermal state (\ref{grandcan}) is the step function at zero
temperature. This means that all the particles occupy the eigenstates
of the hamiltonian with energy smaller than the chemical potential. At
zero temperature the chemical potential is changed only if an
additional particle occupies the first empty eigenstate. Thus, the
smallest change of the chemical potential is the energy difference
between the two eigenstates, and the new thermal state is orthogonal
to the original one. This implies a non-continuous change of the
density matrix, and the state is not differentiable with respect to
the chemical potential. 

We now consider the estimation of the chemical potential, following the
theory for non-differentiable models. Since the temperature should be set to
zero, in order to have a non-differentiable thermal state, we focus on the
single parameter estimation of the chemical potential. The starting point is
to define a finite change of the chemical potential, say $\delta$, and the
finite ratio 

\begin{equation}
\Delta_\delta\rho_{\beta,\mu}=\frac{\rho_{\beta,\mu+\delta}-\rho_{\beta,\mu}}{\delta}.
\end{equation}

\noindent
This ratio plays the role played by the derivative in the Fisher matrix. The symmetric logarithmic derivatives are generalized by the operators $L_{\mu,\delta}$, defined by

\begin{equation}
\Delta_\delta\rho_{\beta,\mu}=\frac{1}{2}\{\rho_{\beta,\mu},L_{\mu,\delta}\},
\end{equation}

\noindent
and the Fisher information is replaced by the following quantity

\begin{equation}
J_{\mu,\delta}={\rm tr}(\rho_{\beta,\mu}L_{\mu,\delta}^2)={\rm tr}(\Delta_\delta\rho_{\beta,\mu}L_{\mu,\delta}).
\end{equation}

\noindent
Finally, the quantum Hammerseley-Chapman-Robbins-Kshirsagar bound reads

\begin{equation} \label{HCRK}
{\rm var}(\mu)\geqslant\frac{1}{J_{\mu,\delta}},
\end{equation}

\noindent
that corresponds to the quantum Cram\'er-Rao bound if $\delta\to 0$ and the
density matrix is differentiable. The inequality (\ref{HCRK}) is sharp, and
the measurement that provides an estimation of $\mu$ with the minimum
uncertainty is a projective measurement on the eigenbasis of
$L_{\mu,\delta}$, i.e. a measurement of the energy and the number of
particles.

If the Fermi energy is degenerate, the thermal state at zero temperature $\rho_{\beta=\infty,\mu}$ is an equally weighted mixture of several pure states. Denote $g$ its degeneracy. The explicit computation in the eigenbasis of the thermal state gives the following result 

\begin{equation}
J_{\mu,\delta}=\frac{g}{\delta^2}.
\end{equation}

At T=0, if the chemical potential is directly tuned, the smallest change
$\delta$ is the energy spacing. The finite parallelepiped-shaped
quantization volumes of homogeneous two- and
three-dimensional gases discussed here break the rotational symmetry. This has some
consequences on the energy spacing. Consider isotropic volumes
$L_x=L_y=L_z=L$ and Fermi momentum almost parallel to one quantization axis
or plane, i.e. only few excitations in at least one direction, say
$n_x\sim{\cal O}(1)$. The distance between the corresponding Fermi energy
and the nearest level is ${\cal
  O}\left(\frac{\hslash^2}{mV_d^{2/d}}\right)$. On the other hand, if the
Fermi energy is characterized by momenta with large wave numbers $n_x\sim
n_y\sim n_z\sim{\cal O}(\eta L)$ with some constant $\eta$, the spacing to
the next enery level is ${\cal
  O}\left(\frac{\hslash^2\eta}{mV_d^{1/d}}\right)$. These two different
scalings give two different regimes for the sensitivity:
$J_{\mu,\delta}\sim{\cal
  O}\left(\frac{gm^2}{\hslash^4\varrho^\frac{4}{d}}(\langle N\rangle_d^{\rm
    hom})^\frac{4}{d}\right)$ and $J_{\mu,\delta}\sim{\cal
  O}\left(\frac{gm^2}{\hslash^4\eta^2\varrho^\frac{2}{d}}(\langle
  N\rangle_d^{\rm hom})^\frac{2}{d}\right)$ respectively, where only the
first scaling is superlinear when the density is finite and $d>1$. Notice that the
Fermi energies which provide the faster scaling of the sensitivity are rare,
because only a few wave number satisfy $n_x\sim{\cal O}(1)$. 

This difference originates in the above mentioned break of the rotational
symmetry which splits some energy levels. Indeed, if the quantization volume
is finite and spherical, the eigenenergies $\frac{p^2}{2m}$ depend only on
one quantum number, i.e. the quantized modulus of the momentum $p$ which
scales with the finite radius of the box as $p\sim{\cal O}(\hslash/R)={\cal
  O}(\hslash/V_d^{1/d})$ \cite{Messiah}. Thus, given the modulus of the
Fermi momentum $k_F$, the energy spacing and the sensitivity are 
\begin{equation} \label{spherical.box}
\delta\sim{\cal O}\left(\frac{\hslash k_F}{mV_d^\frac{1}{d}}\right), \qquad J_{\mu,\delta}\sim{\cal O}\left(\frac{gm^2}{\hslash^2 k_F^2\varrho^\frac{2}{d}}(\langle N\rangle_d^{\rm hom})^\frac{2}{d}\right).
\end{equation}
If the density $\varrho=\langle N\rangle_d^{\rm hom}/V_d$ is fixed, the
scaling of $J_{\mu,\delta}$ with the average number of particles is
sub-linear in three dimensions, linear in two dimensions and superlinear in
one dimension. Note that the square and the spherical quantization volume coincide in one dimension, and thus only the result (\ref{spherical.box}) applies. For harmonically trapped isotropic gases, 

\begin{equation}
\delta=\hslash\Omega_d, \qquad J_{\mu,\delta}=\frac{g}{\hslash^2\tilde\varrho^\frac{2}{d}}(\langle N\rangle_d^{\rm harm})^\frac{2}{d}.
\end{equation}

\noindent
If the density $\tilde\varrho=\langle N\rangle_d^{\rm harm}\Omega_d^d$ is
fixed, $J_{\mu,\delta}$ has the same scaling as in (\ref{spherical.box}).

The absence of superlinear scaling in more than one dimension can be
understood with the presence of high degeneracy in the energy eigenspaces
which becomes continuous in the limit of infinite volume. Indeed, the energy
depends only on the modulus of $\bf p$ in (\ref{cont-hom}) or of $\bf x$ in
(\ref{cont-harm}). Therefore, when the chemical potential is changed, the
new particles or holes are spread on the entire eigenspace. Since the
optimal measurement is a projective measurement onto the Fock states,
i.e. eigenstates of the total number of particles, the continuous degeneracy
of the Fermi surface in two and three dimensions makes the measurement more
difficult than in one dimension with a two-fold degeneracy.  
To further investigate the role of the continuous degeneracy,
consider anisotropies that break this degeneracy. We now focus on gases in
an anisotropic harmonic potential, for the simplicity of the linear energy
spacing, with frequencies $\omega_x=\omega/\alpha_x$,
$\omega_y=\omega/\alpha_y$, $\omega_z=\omega/\alpha_z$, finite $\omega$, and
$\alpha_x>\alpha_y\geqslant\alpha_z$. The typical energy spacing and the
corresponding sensitivity for large anisoptropies $\alpha_x\gg\alpha_{y,z}$
are given by 
\begin{equation}
\delta=\hslash\frac{\omega_x}{\alpha_x}, \qquad J_{\mu,\delta}=\frac{g\alpha_x^2}{\hslash^2\omega^2}.
\end{equation}
If $\alpha_x=\alpha_y^n=\alpha_z^n$, the density is $\tilde\varrho=\langle N\rangle_3^{\rm harm}\omega^3\alpha_x^{-1-2/n}$, and the inverse sensitivity
\begin{equation}
J_{\mu,\delta}=\frac{g \, \omega^{4\frac{n-1}{n+2}}}{\hslash^2\tilde\varrho^{\frac{2n}{n+2}}}(\langle N\rangle_3^{\rm harm})^{\frac{2n}{n+2}}
\end{equation}
scales superlinearly with the average number of particles for $n>2$.  The
fastest possible scaling is quadratic and achieved for $n\to\infty$.

A different situation occurs if one tunes a continuous parameter, e.g. the
voltage or any potential, that leads to jumps of the chemical potential. In this case, $\delta$ can be arbitrary small, for instance one can increase the voltage
by an arbitrary small amount $\delta$, until the overall change of the total
chemical potential equals the energy spacing. At this point, the density
matrix suddenly changes into an orthogonal state, so that the change in the
chemical potential is detected with high accuracy. 

\section{Second order coherence function and the variance of the particle number}
\label{g2}

In this appendix, we prove the relation between the second order coherence
function and the variance of the particle number for homogeneous
one-dimensional gases. Consider the field operator $\Psi^\dag(x)$, namely
the creation operator of a particles localized in position $x$. Field
operators of bosonic particles satisfy commutation relations
$[\Psi(x),\Psi^\dagger(x')]=\delta(x-x')$, while fermionic field operators
satisfy 
anti-commutation relations $\{\Psi(x),\Psi^\dagger(x')\}=\delta(x-x')$. The second
order coherence function is defined as

\begin{equation}
g^{(2)}(x,x')=\frac{\langle\Psi^\dag(x)\Psi^\dag(x')\Psi(x')\Psi(x)\rangle}{\langle\Psi^\dag(x)\Psi(x)\rangle\langle\Psi^\dag(x')\Psi(x')\rangle}.
\end{equation}

\noindent
The total number operator is $N=\int dx \Psi^\dag(x)\Psi(x)$. Therefore, with the help of the (anti)commutation relations, the variance of the total number of particles can be written as

\begin{eqnarray}
\Delta^2 N & = & \int dx dx'\left(\langle \Psi^\dag(x)\Psi(x)\Psi^\dag(x')\Psi(x')\rangle-\langle\Psi^\dag(x)\Psi(x)\rangle\langle\Psi^\dag(x')\Psi(x')\rangle\right) \nonumber \\
& = & \langle N\rangle+\int dxdx' \langle\Psi^\dag(x)\Psi(x)\rangle\langle\Psi^\dag(x')\Psi(x')\rangle\left(g^{(2)}(x,x')-1\right).
\end{eqnarray}

\noindent
The result is the same for bosons and fermions because two exchanges of field operators are required. The hamiltonian of homogeneous gases commutes with the total momentum, and statistical averages depend only on relative distances. Thus, $g^{(2)}(x,x')=g^{(2)}(|x-x'|)$ and $\langle\Psi^\dag(x)\Psi(x)\rangle=\int dx'\langle\Psi^\dag(x')\Psi(x')\rangle/L_x=\varrho$. Exploiting these properties and the change of variables $(x,x')\to(r=x-x',R=(x+x')/2)$ in the previous integral, we get

\begin{equation} \label{var-g2}
\Delta^2 N=\langle N\rangle+2\varrho^2\int_{0}^{L_x}dr(L_x-r)\left(g^{(2)}(r)-1\right).
\end{equation}

\noindent
Notice that this equation differs form that presented in \cite{Yukalov2005},
which gives the leading contribution for large $L_x$.





\end{document}